\documentclass[twocolumn,epjc3]{svjour3}
\usepackage{graphics}
\usepackage{microtype}
\usepackage{bbm}
\usepackage{amsmath}
\usepackage{mathtools}
\usepackage{caption}
\usepackage{subcaption}
\usepackage{amssymb}
\usepackage{mathrsfs}       
\usepackage[pdftex]{graphicx}
\usepackage{pgfplots}
\pgfplotsset{width=10cm,compat=1.9}
\usepackage{xargs}       
\usepackage{wasysym}
\usepackage{engrec}
\usepackage{enumerate}
\usepackage{appendix}
\usepackage{empheq}
\usepackage{float}
\usepackage{stmaryrd}
\usepackage[parfill]{parskip}
\usepackage[pdftex,breaklinks]{hyperref}
\usepackage{titletoc}
\usepackage{makecell}
\usepackage{lscape}
\usepackage{algorithm}
\usepackage{algorithmicx}
\usepackage{tensor}
\usepackage{algpseudocode}
\usepackage{blkarray}
\usepackage{multirow}
\usepackage{bigstrut}
\usepackage{booktabs}
\usepackage{siunitx}
\usepackage{mathtools}
\usetikzlibrary{plotmarks}
\usepackage{comment}

\newcommand{\ket}[1]{|#1\rangle}

\newenvironment{customlegend}[1][]{%
    \begingroup
    \csname pgfplots@init@cleared@structures\endcsname
    \pgfplotsset{#1}%
}{%
    \csname pgfplots@createlegend\endcsname
    \endgroup
}%
\def\addlegendimage{\csname pgfplots@addlegendimage\endcsname}

\newcommand{\emax}{e_{\mathrm{max}}}

\definecolor{mygreen}{RGB}{21, 158, 10}
\begin{document}
\title{Deformed natural orbitals for \emph{ab initio} calculations}

\author{A. Scalesi\thanksref{ad:saclay} 
\and T. Duguet\thanksref{ad:saclay,ad:kul} 
\and M. Frosini\thanksref{ad:des}
\and V. Som\`a\thanksref{ad:saclay}
}

\institute{
\label{ad:saclay}
IRFU, CEA, Universit\'e Paris-Saclay, 91191 Gif-sur-Yvette, France 
\and
\label{ad:kul}
KU Leuven, Department of Physics and Astronomy, Instituut voor Kern- en Stralingsfysica, 3001 Leuven, Belgium 
\and
\label{ad:des}
CEA, DES, IRESNE, DER, SPRC, LEPh,
13108 Saint-Paul-lez-Durance, France
}

\maketitle
%
% The correct dates will be entered by Springer
%
\begin{abstract}
The rapid development of \emph{ab initio} nuclear structure methods towards doubly open-shell nuclei, heavy nuclei and greater accuracy occurs at the price of evermore increased computational costs, especially RAM and CPU time.
While most of the numerical simulations are carried out by expanding relevant operators and wave functions on the spherical harmonic oscillator basis, alternative one-body bases offering advantages in terms of computational efficiency have recently been investigated.
In particular, the so-called natural basis used in combination with symmetry-conserving methods applicable to doubly closed-shell nuclei has proven beneficial in this respect. 
The present work examines the performance of the natural basis in the context of \emph{symmetry-breaking} many-body calculations enabling the description of superfluid and deformed open-shell nuclei at polynomial cost with system's size.
First, it is demonstrated that the advantage observed for closed-shell nuclei carries over to  open-shell ones.
A detailed investigation of natural-orbital wave functions provides useful insight to support this finding and to explain the superiority of the natural basis over alternative ones.
Second, it is shown that the use of natural orbitals combined with importance-truncation techniques leads to an even greater gain in terms of computational costs. 
The present results pave the way for the systematic use of natural-orbital bases in future implementations of non-perturbative many-body methods.
\end{abstract}

\section{Introduction}
\label{introduction}

To these days, \emph{ab initio} methods~\cite{WhatIsAbInitio} already provide a reliable description of a large set of atomic nuclei.
Such methods aim at approximating the solutions of the many-body Schr\"odinger equation in a systematic way starting from inter-nucleon interactions rooted in the gauge theory of the strong force, i.e. quantum-chromodynamics, via the use of chiral effective field theory~\cite{Hergert20,Hammer2020a}.
This is typically achieved by representing relevant quantities (wave functions, operators, density matrices, etc.) on a basis of the A-body Hilbert space $\mathcal{H}_\text{A}$, itself obtained as the tensor-product of bases of the one-body Hilbert space $\mathcal{H}_1$.
Because of finite computational resources, the infinite-dimensional basis of $\mathcal{H}_\text{A}$ has to be truncated to perform practical calculations, either directly or via a truncation of the underlying one-body basis. Eventually, the size of the truncated basis impacts both the cost of handling (in terms of storage and RAM) the Hamiltonian tensors constituting the input of a given simulation and the cost of solving the many-body Schr\"odinger equation (in terms of storage, RAM and CPU time) to determine the many-body tensors (wave function, density matrices...) constituting the output. Clearly, the accuracy of a calculation depends on the appropriateness of the chosen basis truncation, i.e. on the \emph{model-space} truncation, which itself depends on the {\it characteristics} of the employed basis. While attaining a suitable error\footnote{The model-space uncertainty must be similar to or smaller than the other sources of error in the many-body calculation.} does not generally constitute a difficulty in light nuclei, it may become challenging in medium-mass nuclei and ultimately limits the application of state-of-the-art techniques to heavy systems (see, e.g.,~\cite{Hu22}), especially when aiming at doubly open-shell nuclei and/or at solving the many-body Schr\"odinger equation with sub-percent accuracy (see, e.g.,~\cite{Hu24}); see Ref.~\cite{frosini2024tensor} for a detailed discussion.

Several strategies are currently pursued to alleviate the computational cost of many-body calculations (at fixed accuracy). First, importance truncation (IT)~\cite{Roth:2009eu,TICHAI2019,IMSRG_IT,Porro:2021rw} and tensor factorisation (TF)~\cite{Tichai:2018eem,TICHAI2019,Tichai:2021rtv,Tichai:2023dda,frosini2024tensor} techniques aim at reducing the storage and CPU footprints of input and output tensors while working in a given one-body basis of choice, typically the eigenbasis of the one-body spherical harmonic oscillator (sHO) Hamiltonian. A second approach, the one followed in the present work, consists of optimising in a first step the {\it nature} of the one-body basis in order to reach faster convergence with respect to the cardinality of that basis. Eventually, the two strategies can be combined to push the limits of state-of-the-art calculations.

The use of the sHO one-body basis offers several advantages in practical applications. First, the analytical knowledge of the sHO single-particle wave functions allows for a convenient representation of the operators at play in the problem~\cite{Moshinsky}. Second, truncating appropriately the many-body basis of Slater determinants built from sHO one-body states authorizes an exact\footnote{Performing the model-space truncation at the level of the sHO one-body basis as  done in so-called expansion many-body methods, an {\it effective} centre-of-mass factorization is achieved~\cite{Hagen09}.} factorisation of the centre-of-mass wave-function~\cite{Caprio20}. On the other hand, the fact that sHO states decay at long distances as a Gaussian function rather than as an exponential one makes difficult in practice to represent loosely-bound many-body states or to bridge to nuclear reactions.

While the optimisation of one-body basis states is central in electronic structure calculations~\cite{Davidson72,HelgakerBook}, the use of alternatives to the sHO basis has received limited attention in nuclear physics, with only a few exceptions~\cite{CoulombSturmian,puddu,Negoita10,BulgacForbes}. In the present case, one is not only interested in exploring alternative bases that would be given {\it a priori} but rather to employ a {\it nucleus-dependent} basis that is informed of the characteristics of the system under consideration; i.e. a basis that reflects the bulk of many-body correlations in order to best accelerate the convergence (with respect to the one-body basis size) of a subsequent high-accuracy calculation of those correlations. 

A successful choice in this respect is provided by the natural (NAT) orbital basis~\cite{benderNAT,DavidsonNAT,JeffreyHayNAT,Robin16,Robin17,Robin21} obtained by diagonalising the one-body density matrix of the correlated state under consideration. In particular, a faster convergence of ground-state observables has been found both in exact diagonalisation techniques~\cite{Fasano22,Tichai19} applicable to light nuclei, and in calculations of doubly closed-shell nuclei based on symmetry-conserving expansion methods~\cite{Hoppe21}.
While natural orbitals have also been recently employed in deformed coupled-cluster calculations~\cite{Novario20}, a detailed study of their performance in \emph{symmetry-breaking} expansion methods applicable to all nuclei is currently missing.

The first goal of the present article is thus to investigate the use of the NAT basis in expansion methods based on superfluid and deformed reference states dedicated to singly and doubly open-shell nuclei. Because symmetry-breaking methods necessitate a much larger number of one-body basis states than symmetry conserving ones, their optimisation is even more compelling. To do so, deformed Bogoliubov many-body perturbation theory (dBMBPT)~\cite{Frosini:2021tuj,frosini2024tensor} at second or third order is employed to both generate the one-body density matrix from which the NAT basis is extracted and compute ground-state energy out of which the accelerated convergence is characterised. The second objective of this work is to compare the benefits obtained using the NAT basis and IT techniques before combining both tools.

The paper is organized as follows. 
Section~\ref{formalism} introduces the main theoretical and computational ingredients, including details on the extraction of the natural basis. Section~\ref{standardNAT} compares the performance of the NAT and sHO bases in open-shell nuclei.
Section~\ref{investigation} examines possible alternatives to the NAT basis and provides further insight by analysing the behaviour of the associated single-particle wave functions. In Sec.~\ref{nat-it} the NAT basis is compared to IT and combined with it.  Finally, Sec.~\ref{conclusions} summarizes the main conclusions and discusses possible future developments.

\section{Formalism and computational setting}
\label{formalism}

\subsection{Many-body method}

The goal is to extract {\it approximate} natural orbitals from a many-body state informed of bulk of many-body correlations via a calculation that is significantly less costly than the one of interest. A low-order deformed Bogoliubov many-body perturbation theory~\cite{Frosini:2021tuj,Scalesi24b} calculation based on a deformed Hartree Fock Bogoliubov (dHFB) unperturbed state is ideally suited to do so across a significant part of the nuclear chart, independently of the closed-, singly open- and doubly open-shell character of the nucleus under consideration. The method used to generate the NAT basis will be indicated in square brackets, e.g. ``NAT[dBMBPT(2)]" denotes the NAT basis obtained from a second-order dBMBPT density matrix. Whenever the unperturbed state is actually unpaired, ``dHF" or ``dMBPT" can be used to label the calculation in use. 

While the objective is to eventually perform non-perturbative calculations of open-shell nuclei\footnote{One typically has in mind to perform deformed coupled cluster (dCC)~\cite{Novario20,Hagen:2022tqp}, in-medium similarity renormalization group (dIMSRG)~\cite{Yuan:2022gyo} or Dyson self-consistent Green's function (dDSCGF)~\cite{dDSCGF} calculations based on a deformed reference state.}, dBMBPT is also used in the present paper to validate the acceleration offered by the use of the NAT basis. Computations in this manuscript are carried out with the {\tt{PAN@CEA}}~\cite{pan@cea} numerical code, which implements dHFB and dBMBPT(2,3) equations. 

\subsection{Spherical harmonic oscillator basis}

Input Hamiltonian tensors are presently represented using the sHO one-body basis. sHO states are characterised by the set of quantum numbers 
\begin{align}
\alpha \equiv (n_\alpha, \pi_\alpha, j_\alpha, m_\alpha, t_\alpha) \, , \label{set1}
\end{align}
where $n_\alpha$ denotes the principal quantum number, $\pi_\alpha=(-1)^{l_\alpha}$ the parity, with $l_\alpha$ being the orbital angular momentum, $j_\alpha$ the total angular momentum whereas $m_\alpha$ and $t_\alpha$ represent the projection of the total angular momentum and of the isospin along the quantisation axis, respectively. The dimension $n_B$ of the basis, i.e. the range of the index $\alpha$, is set by selecting states according to $0 \leq e_\alpha \leq e_\text{max}$ with $e_\alpha \equiv 2n_\alpha +l_\alpha$. The values of $n_B$ corresponding to $2\leq e_\text{max} \leq 14$ are displayed in Tab.~\ref{tab:emax_comp}.

The spatial extension of sHO states can be tuned via the choice of the frequency $\hbar \omega$ of the harmonic oscillator potential. Even if ab initio calculations become independent of that choice when a large enough $e_\text{max}$ is employed, a particular value of $\hbar \omega$ can help optimize the convergence of the calculation for a given $e_\text{max}$. For any choice of the oscillator frequency, though, all sHO basis states display a wrong asymptotic behavior at large distances, i.e. they fall off as Gaussian functions, which makes it difficult to reproduce the exponential tail of the one-nucleon density distribution.

\subsection{Deformed quasi-particle basis}

The dHF(B) unperturbed state $| \Phi \rangle$ at play in d(B)MBPT, dCC and dDSCGF breaks rotational invariance. Consequently, the quasi-particle basis\footnote{In case the unperturbed state is a deformed Hartree-Fock Slater determinant, the quasi-particle basis relates trivially to the deformed Hartree-Fock single-particle basis. In case the unperturbed state is a deformed Hartree-Fock-Bogoliubov state, it corresponds to the actual deformed Bogoliubov quasi-particle basis.} labelling the many-body tensors at play in the method of interest~\cite{frosini2024tensor} is characterized by the set of quantum numbers
\begin{align}
\alpha \equiv (N_\alpha, \pi_\alpha, m_\alpha, t_\alpha) \, , \label{set2}
\end{align}
where $N_\alpha$ denotes a novel principal quantum number. While $\pi_\alpha$ and $m_\alpha$ remain good quantum numbers\footnote{While it can be further broken~\cite{frosini2024tensor}, the rotational symmetry around the $z$ axis is presently conserved, i.e. the unperturbed state remains axially symmetric.}, it is not anymore the case for $j_\alpha$ and $l_\alpha$.

\begin{table}[t]
\centering
\begin{tabular}{c|c|c}
$e_\text{max}$ & $n_B$\\
\hline
2 &  40 \\
4 & 140 \\
6 & 336 \\
8 & 660 \\
10 & 1144 \\
12 & 1820 \\
14 & 2720 \\
\end{tabular}
\caption{Number of sHO one-body basis states $n_B$ for a given $e_\text{max}$ truncation. For each isospin, the number of sHO one-body basis states is $n_B^P=n_B^N\equiv n_B/2$.}
\label{tab:emax_comp}
\end{table}

\subsection{Deformed natural orbital basis}
\label{natural}

\subsubsection{Definition}

Given the dHFB unperturbed vacuum $| \Phi \rangle$ at hand, the dBMBPT($p$=2,3) many-body state reads as
\begin{align}    
| \Psi^{(p)} \rangle =& | \Phi \rangle \nonumber \\ 
&+ \frac{1}{(1)!}  \sum_{k_1k_2} C^{2 0}_{k_1k_2}(p)| \Phi^{k_1k_2} \rangle  \nonumber \\ 
&+ \frac{1}{(4)!} \sum_{k_1\ldots k_4} C^{4 0}_{k_1k_2k_3k_4}(p)| \Phi^{k_1k_2k_3k_4} \rangle \nonumber \\ 
&+ \frac{1}{(6)!} \sum_{k_1\ldots k_6} C^{6 0}_{k_1k_2k_3k_4k_5k_6}(p)| \Phi^{k_1k_2k_3k_4k_5k_6} \rangle \, ,
\label{eq:stateBMBPT2}
\end{align}
where $| \Phi^{k_1\cdots k_{2q}} \rangle$ denote elementary excitations obtained via the action of $2q$ quasi-particle creation operators on $| \Phi \rangle$. The presence of single, double and triple excitations signals that the bulk of dynamical correlations is captured by $| \Psi^{(p)} \rangle$ in addition to the static correlations resummed into $| \Phi \rangle$ through the breaking of $U(1)$ and $SU(2)$ symmetries. The explicit expressions of the second-order coefficients $C^{2 0}_{k_1k_2}(2)$ and $C^{4 0}_{k_1k_2k_3k_4}(2)$  can be found in Ref.~\cite{Frosini:2021tuj}. Since $| \Psi^{(2)} \rangle$ does not contain triple excitations, one has $C^{6 0}_{k_1k_2k_3k_4k_5k_6}(2)=0$.

Having the dBMBPT($p$) state $| \Psi^{(p)} \rangle$ at hand, the associated normal one-body density matrix is computed in the sHO basis as\footnote{The perturbative order employed here to label the density matrix is unconventional. Indeed, it corresponds to the truncation order of the many-body state rather than of the density matrix itself. This leads us to keeping terms that are omitted in the conventional definition used in, e.g.,  Ref.~\cite{Tichai:2019to} that omits contributions beyond second order in Eq.~\eqref{eqn:onebodydensop}. This is of little importance here given that the density matrix is only used to produce an auxiliary one-body basis of interest whose characteristics are essentially insensitive to the additional terms presently included.}
\begin{align}
\rho_{\alpha\beta}^{(p)} &\equiv \frac{\langle \Psi^{(p)} | a^\dagger_\beta a_\alpha | \Psi^{(p)} \rangle}{\langle \Psi^{(p)} |  \Psi^{(p)} \rangle} \nonumber \\
&\equiv \delta_{m_\alpha m_\beta}\delta_{\pi_\alpha \pi_\beta}\delta_{t_\alpha t_\beta}\, \rho_{n_\alpha j_\alpha n_\beta j_\beta}^{[m\pi t]_\alpha \, (p)} \, ,
\label{eqn:onebodydensop}
\end{align}
such that it is hermitian and positive semi-definite. For reference, the explicit expression of $\rho^{(2)}$ can be found in Ref.~\cite{Frosini:2021tuj}.

The one-body density matrix is diagonalized according to
\begin{equation}
\sum_{\beta} \rho_{\alpha\beta}^{(p)} C_{\beta \gamma}^{(p)} = \lambda_\gamma^{(p)} C_{\alpha\gamma}^{(p)} \, . \label{diagorho}
\end{equation}

The eigenstates of $\rho^{(p)}$ are nothing but the deformed NAT[dBMBPT($p$)] basis states whose eigenvalues $\lambda_\gamma^{(p)}$ denote their average occupation in $| \Psi^{(p)} \rangle$. In Eq.~\eqref{diagorho}, the eigenvectors provide the unitary transformation between the sHO basis $\{| \varphi_\alpha \rangle; \alpha = 1,\ldots, n_B\}$ and the NAT[dBMBPT($p$)] basis  $\{| \phi_\gamma^{(p)} \rangle; \gamma = 1,\ldots, n_B\}$
\begin{equation}
C_{\alpha\gamma}^{(p)} \equiv \langle \varphi_\alpha | \phi_\gamma^{(p)} \rangle \, .
\end{equation}

\subsubsection{Algorithm}

In practice, the density matrix is diagonalized in each separate $[m \pi t]_\alpha$ block such that the transformation between the two bases reads as
\begin{equation}
\begin{split}
\ket{N_\alpha m_\alpha \pi_\alpha t_\alpha}_{\text{NAT[dBMBPT($p$)]}} 
& = \sum_{n_\alpha j_\alpha} C^{[m\pi t]_\alpha \, (p)}_{n_\alpha j_\alpha N_\alpha} \\
& \times
\ket{n_\alpha \pi_\alpha j_\alpha m_\alpha t_\alpha}_{\rm sHO} \, .
\end{split}
\label{eqn:NATcoeff}
\end{equation}

In each $[m \pi t]_\alpha$ block, the principal quantum number $N_\alpha$ are arranged according to the decreasing occupation  ($\lambda_1^{(p)} \geq \lambda_2^{(p)} \geq \ldots$) of the NAT states. Based on this ordering, an \emph{effective} $\tilde{e}_\text{max}$ parameter\footnote{In any given calculation, $\tilde{e}_\text{max}$ is necessarily smaller than or equal to the $e_\text{max}$ value originally used to compute the density matrix.} is defined such that the number of NAT states retained is the same as for the sHO truncated according to $e_\text{max}=\tilde{e}_\text{max}$
. 

Truncating the  NAT[dBMBPT($p$)] basis according to $\tilde{e}_\text{max}$, the matrix elements of the one-body kinetic energy and of the two-body interaction\footnote{The two-body part of the Hamiltonian includes the genuine two-body interaction, the two-body part of the center-of-mass correction and the two-body part of the rank-reduced three-body interaction.} initially expressed in the sHO basis are transformed into the deformed NAT basis using Eq.~\eqref{eqn:NATcoeff}. Based on this (reduced) set of matrix elements, a dHFB state is recomputed and the expansion method of choice is performed on top of it.

\begin{figure}
\includegraphics[width=1.0\columnwidth]{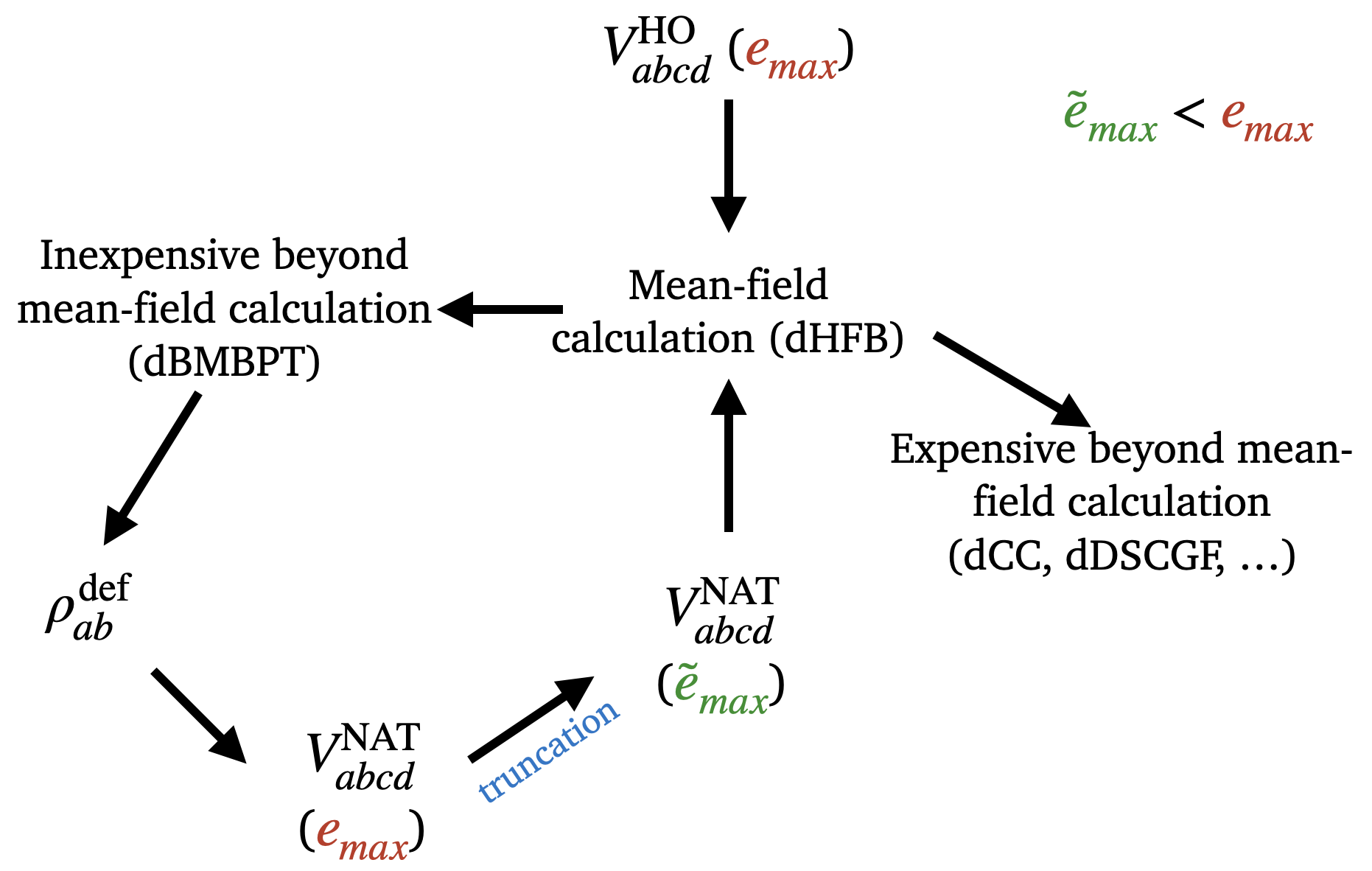}
\caption{\label{fig:NAT_workflow} Workflow of the generation of the NAT basis via a dBMBPT calculation, leading to a redefinition of the interaction matrix elements for a subsequent many-body calculation.}
\end{figure}

\subsubsection{Basic properties}
\label{propNO}

Natural orbitals possess the key extremum property~\cite{lowdin60a}
\begin{equation}
\sum_{\gamma = 1}^{r\leq n_B} \lambda_\gamma^{(p)}  \leq \sum_{k = (1)}^{(r)\leq n_B} \rho_{kk}^{(p)}
\end{equation}
where the diagonal elements $\rho_{kk}$ can here be taken in {\it any} one-body basis and where in the sum over $k= (1),\ldots,(r)$ any set of $r$ indices among the $n_B$ ones can be selected. This property expresses the fact that the occupations fall as quickly as possible in the natural basis such that any subset gathering the most occupied orbitals is indeed maximally occupied. In turn, this property implies that the expansion of the many-body state on the set of Slater determinants built out of NAT states displays optimal convergence properties~\cite{lowdin60a}. Thus, one expects the above property to translate into the fact that the use of the NAT basis optimally accelerates the convergence of a given many-body expansion method with respect to the $\tilde{e}_\text{max}$ truncation. If so, extracting natural orbitals from an inexpensive dBMBPT($p$) calculation with a large enough $e_\text{max}$ may authorize in a second step to converge a more expensive calculation for $\tilde{e}_\text{max}<e_\text{max}$. 

A second interesting property relates to the asymptotic behavior of NAT states that can be inferred from the local nucleon density distribution given by 
\begin{align}
\rho^{(p)}(\vec{r}) = \sum_{\gamma} \lambda^{(p)}_{\gamma} |\phi_\gamma^{(p)}(\vec{r})|^{2} \, .
\label{localdens}
\end{align}
Due to the finite-range of nuclear forces, the long-distance behaviour of the one-nucleon density distribution is given by~\cite{Rotival:2007hp}
\begin{align}
\rho^{(p)}(\vec{r}) &\underset{r\rightarrow+\infty}{\longrightarrow} \frac{e^{-2\kappa_0 \,r}}{(\kappa_0\,r)^2} \, ,
\label{localdensasympt}
\end{align}
with $ \kappa_0=\sqrt{-2m\epsilon_0/\hbar^2}$ and where $\epsilon_0=\left(E_0^{N}-E_0^{N-1}\right)$ is minus the one-nucleon separation energy to reach the ground state of the system with one less nucleon. Because $\rho^{(p)}(\vec{r})$ decays exponentially with a rate set by the one-nucleon removal energy, and because all contributions in the right-hand side of Eq.~\eqref{localdens} are strictly positive\footnote{As soon as $| \Psi^{(p)} \rangle $ does not restrict to a Slater determinant, all eigenvalues of the one-body density matrix are strictly positive in principle. In practice of course, several eigenvalues can be identified as a numerical zero.}, all natural orbital wave-functions are localized and decay faster than $\rho^{(p)}(\vec{r})$ at long distances.

This is to be compared to the case where the many-body state reduces to a single, e.g. dHF, Slater determinant. In this case natural orbitals are nothing but HF single-particle states and Eq.~\eqref{localdens} must be replaced by
\begin{align}
\rho^{(\text{dHF})}(\vec{r}) = \sum_{\alpha \in \text{occ.}} |\psi^{(\text{dHF})}_\alpha(\vec{r})|^{2} \, .
\label{localdensHF}
\end{align}
It follows that the above property only applies to the occupied ($\lambda^{(\text{dHF})}_{\alpha}=1$) single-particle states in the Slater determinant, while all unoccupied ($\lambda^{(\text{dHF})}_{\alpha}=0$) single-particle HF states are not constrained to decay exponentially. In fact, the latter actually oscillate to infinite distance as scattering states as soon as the corresponding HF single-particle energy is positive. These characteristics will be useful later on to analyse the results obtained with different one-body bases.

\subsection{Hamiltonian}
\label{hamiltonians}

Two instances of nuclear Hamiltonians generated within the frame of chiral effective field theory ($\chi$EFT) are employed in the present work
\begin{itemize}
\item
${\rm NNLO}_{\rm sat}$ (bare)~\cite{NNLOsat} ;
\item
EM 1.8/2.0~\cite{MagicInt} .
\end{itemize}
While the second Hamiltonian is directly built as a soft representative displaying negligible coupling between low and high (relative) momentum states, the former does display significant coupling to high momenta. In order to transition from the latter to the former and characterize the impact of coupling to high momenta on the accelerated convergence induced by the NAT basis, the ${\rm NNLO}_{\rm sat}$ Hamiltonian is further evolved through a free-space similarity renormalisation group transformation (SRG)~\cite{Bogner:2009bt} in order to decouple low and high momenta. Doing so, down to the momentum scale $\SI{2.4}{fm^{-1}}$ ($\SI{2.0}{fm^{-1}}$), one defines the evolved ${\rm NNLO}_{\rm sat}$ (2.4) (${\rm NNLO}_{\rm sat}$ (2.0)) Hamiltonians.

Reference calculations employ an $e_\text{max}=12$ truncation of the sHO basis. Calculations with the EM 1.8/2.0 Hamiltonian are performed with an oscillator frequency  $\hbar\omega=20$ MeV while results relative to ${\rm NNLO}_{\rm sat}$ (bare) are obtained with $\hbar\omega=18$ MeV (except if specified otherwise). Three-body interaction matrix elements are further truncated to $e_\text{3max}=16< 3 \, e_\text{max}$ before reducing the three-body interaction to an effective two-body one via the rank-reduction  method developed in Ref.~\cite{Frosini:2021tuj} on the basis of the normal one-body density matrix obtained via a spherical HFB calculation. 

Based on this numerical setting the performance of a given one-body basis is characterized by computing the relative error $\Delta E [\%]$ of the dBMBPT ground-state energy obtained for a given $\tilde{e}_\text{max}\leq 12$ truncation with respect to the reference results obtained for $e_\text{max}=12$.

\begin{figure*}
\centering
\includegraphics[width=2.1\columnwidth]{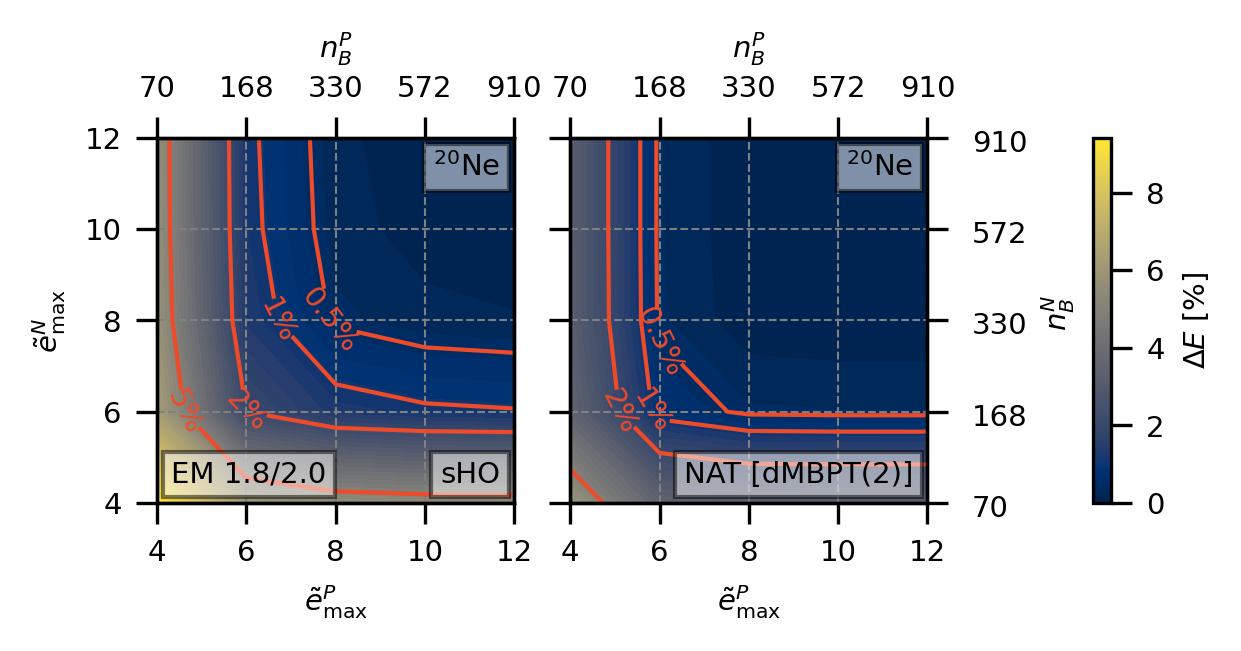}
\caption{\label{fig:Ne20_conv} Convergence of the $^{20}$Ne ground-state energy (relative to the $e_\text{max}=12$ value) computed at the dMBPT(2) level using the sHO (\emph{left panel}) and NAT[dMBPT(2)] (\emph{right panel})) bases.
Results are shown as a function of $\tilde{e}^N_\text{max}$ and $\tilde{e}^P_\text{max}$.
For each basis truncation, the corresponding number of basis states are shown on the upper and right axes (See also Tab.~\ref{tab:emax_comp}.). Calculations are performed with the EM 1.8/2.0 Hamiltonian.}
\end{figure*}

\section{NAT[dBMBPT(2)] basis performance}
\label{standardNAT}

The goal of the present work is to assess the accelerated convergence obtained by using the NAT[dBMBPT(2)] basis relative to the standard HO basis in realistic calculations of doubly open-shell nuclei. Because the NAT basis is isospin dependent, the {\tt{PAN@CEA}}~\cite{pan@cea} code has been extended to handle such a feature. In turn, this makes possible to truncate separately neutron and proton one-body basis states according to the $\tilde{e}^N_{\text{max}}$ and $\tilde{e}^P_{\text{max}}$ parameters.

\subsection{sHO vs NAT[dBMBPT(2)] bases}
\label{HOvsNAT}

The first nucleus under study is $^{20}$Ne, a prolate nucleus recently investigated via various \emph{ab initio} expansion methods~\cite{Novario20,Frosini:2021sxj,Frosini:2021ddm}. The convergence of the ground-state energy computed with the EM 1.8/2.0 Hamiltonian is displayed in Fig.~\ref{fig:Ne20_conv} for both the sHO and the NAT[dBMBPT(2)] bases as a function of $\tilde{e}^N_{\text{max}}$ and $\tilde{e}^P_{\text{max}}$.

The NAT basis is seen to display a faster convergence than the sHO basis, e.g. the 1\% error with respect to the converged ($e_\text{max}=12$) result is reached at $\tilde{e}^N_{\text{max}}=\tilde{e}^P_{\text{max}}=8$ in the sHO basis whereas  $\tilde{e}^N_{\text{max}}=\tilde{e}^P_{\text{max}}=6$ is sufficient in the NAT basis. Because  $^{20}$Ne is a $N=Z$ nucleus, the gain is symmetric with respect to $\tilde{e}^N_{\text{max}}$ and $\tilde{e}^P_{\text{max}}$.

As seen from Tab.~\ref{tab:emax_comp}, such an advantage allows one to work with about half the number $n_B$ of states compared to the sHO basis, which already constitutes a sizeable advantage in terms of storage and CPU time for any expansion method scaling polynomially, i.e. as $n^q_B$, with the one-body basis size. For state-of-the-art high-accuracy methods for which $q=7$ or $8$~\cite{kucharski_coupled-cluster_1992,sun_how_2022-1,danovich11a,Cipollone:2013zma,Barbieri:2021ezv,Heinz:2021xir,He:2024utz,Stroberg:2024cki}, the gain can be very significant.

\begin{figure*}
\centering
\includegraphics[width=2.1\columnwidth]{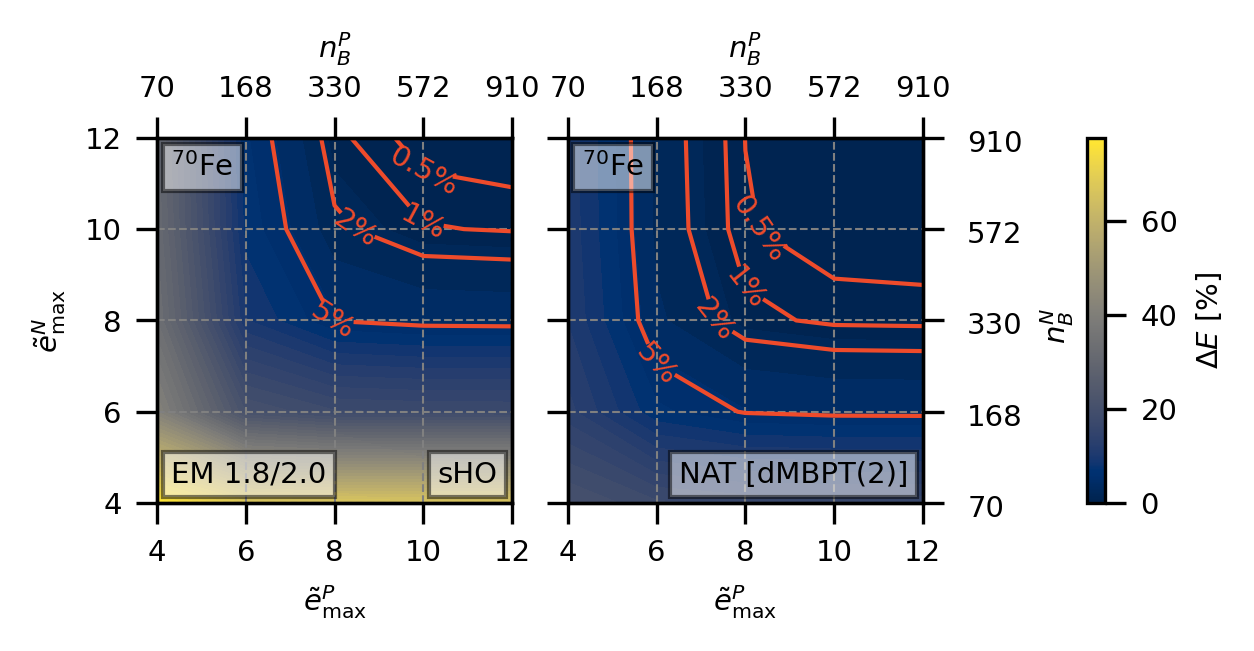}
\caption{\label{fig:Fe70_conv} Same as Fig.~\ref{fig:Ne20_conv} for $^{70}{\rm Fe}$.}
\end{figure*}
\begin{figure*}
\centering
\includegraphics[width=2.1\columnwidth]{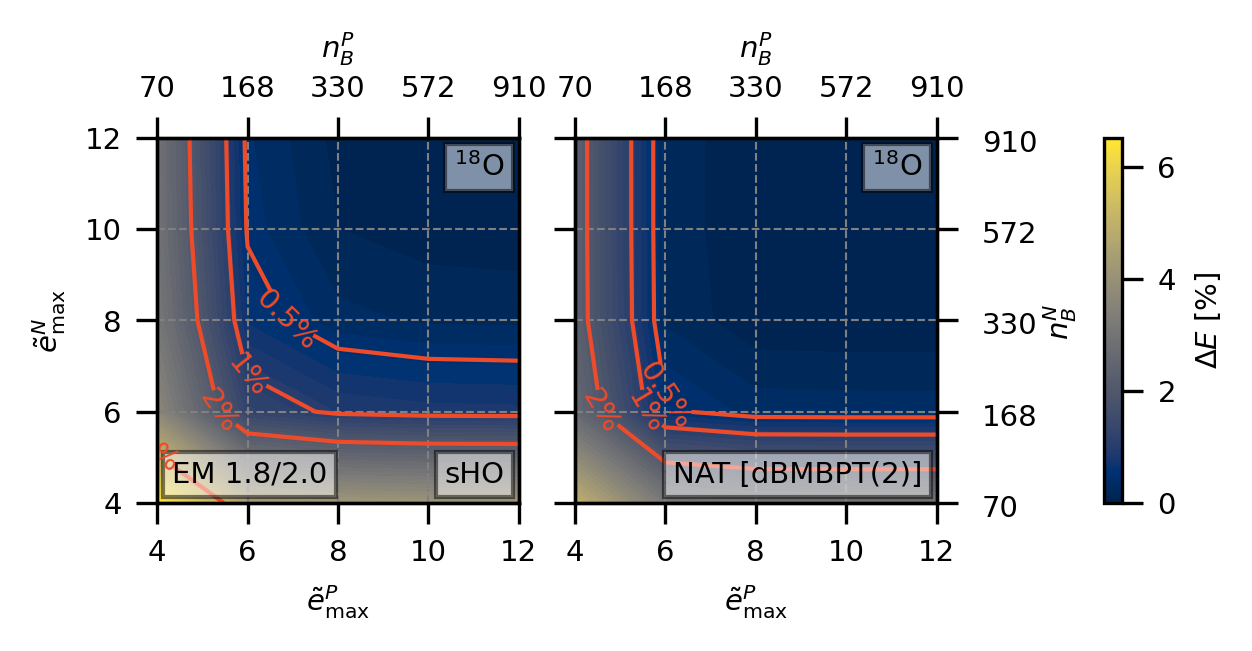}
\caption{\label{fig:O18_conv} Same as Fig.~\ref{fig:Ne20_conv}  for $^{18}{\rm O}$.}
\end{figure*}

Next, the NAT basis is tested on a heavier neutron-rich prolate $^{70}$Fe nucleus; results are shown in Fig.~\ref{fig:Fe70_conv}. The error associated with the sHO basis displays an asymmetric pattern, the energy converging faster with respect to $\tilde{e}^P_{\text{max}}$ than to $\tilde{e}^N_{\text{max}}$. The use of the NAT basis reduces the neutron-proton asymmetry and an advantage analogous to the one obtained for $^{20}$Ne is observed. The fact that the benefit carries over to medium-heavy mass deformed nuclei is encouraging in view of using the NAT basis for the most computationally challenging systems in the future.

Finally, Fig.~\ref{fig:O18_conv} displays results for the singly open-shell, i.e. spherical and superfluid, $^{18}$O nucleus.
For a small, i.e. $0.5\%$ or $1\%$, error, a similar advantage to the one observed $^{20}$Ne and $^{70}$Fe is achieved.

\subsection{Application to dBMBPT(3)}
\label{NATHFB}

\begin{figure*}
\centering
\includegraphics[width=2.1\columnwidth]{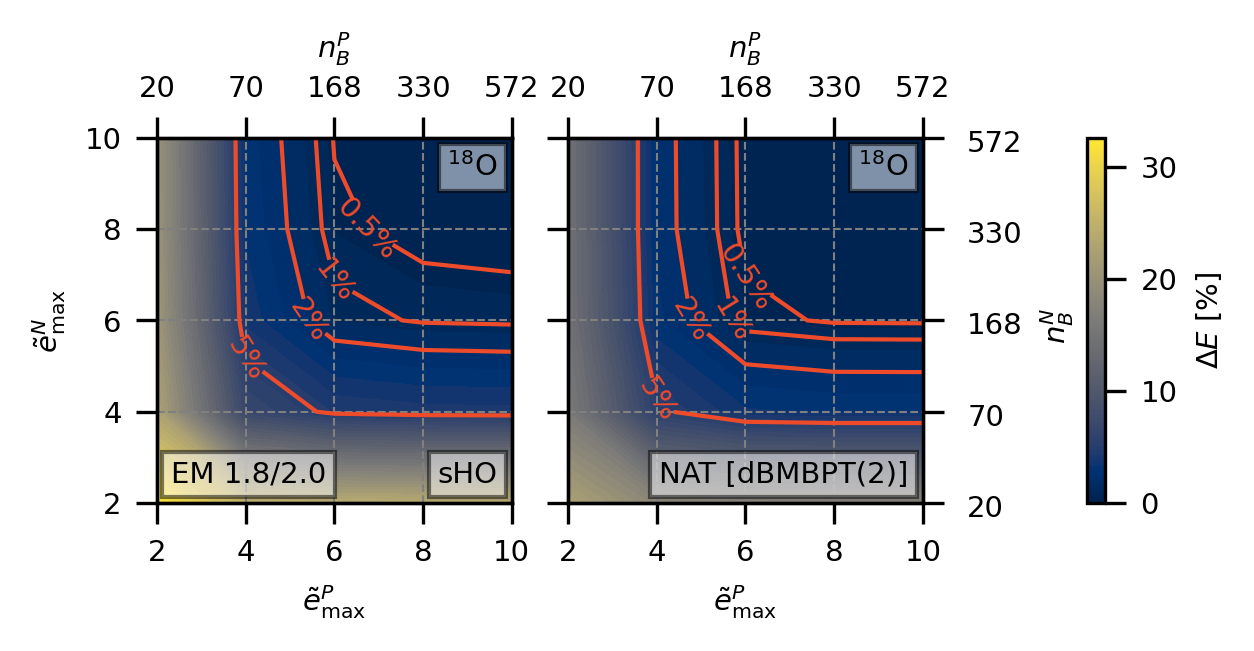}
\caption{\label{fig:O18_BMBPT3} Same as Fig.\ref{fig:O18_conv} but for a dBMBPT(3) calculation and an error computed relatively to the $e_\text{max}=10$ calculation.}
\end{figure*}

Having tested the performance of the NAT[dBMBPT(2)] basis through dBMBPT(2), one can employ a more advanced dBMBPT(3) calculation to further validate the conclusions. The result of such a test, reported in Fig.~\ref{fig:O18_BMBPT3} for $^{18}$O, indeed leads to similar conclusions as for dBMBPT(2)\footnote{Other possibilities, i.e. dBMBPT(2) calculations on top of NAT[dBMBPT(3)] or dBMBPT(3) calculations on top of NAT[dBMBPT(3)] have also been tried and all lead to similar results.}.

\subsection{Resolution-scale dependence}

The correlations encoded in a beyond-mean-field, e.g. dBMBPT(2), density matrix ultimately depend on the input Hamiltonian, and in particular on its resolution scale.
It is thus important to assess the efficiency of the NAT machinery for interactions characterised by different degrees of `softness'.
To this end, the four Hamiltonians introduced in Sec.~\ref{hamiltonians}, spanning a significant range of resolution scales, are now considered.

\begin{figure*}
\centering
\includegraphics[width=2.\columnwidth]{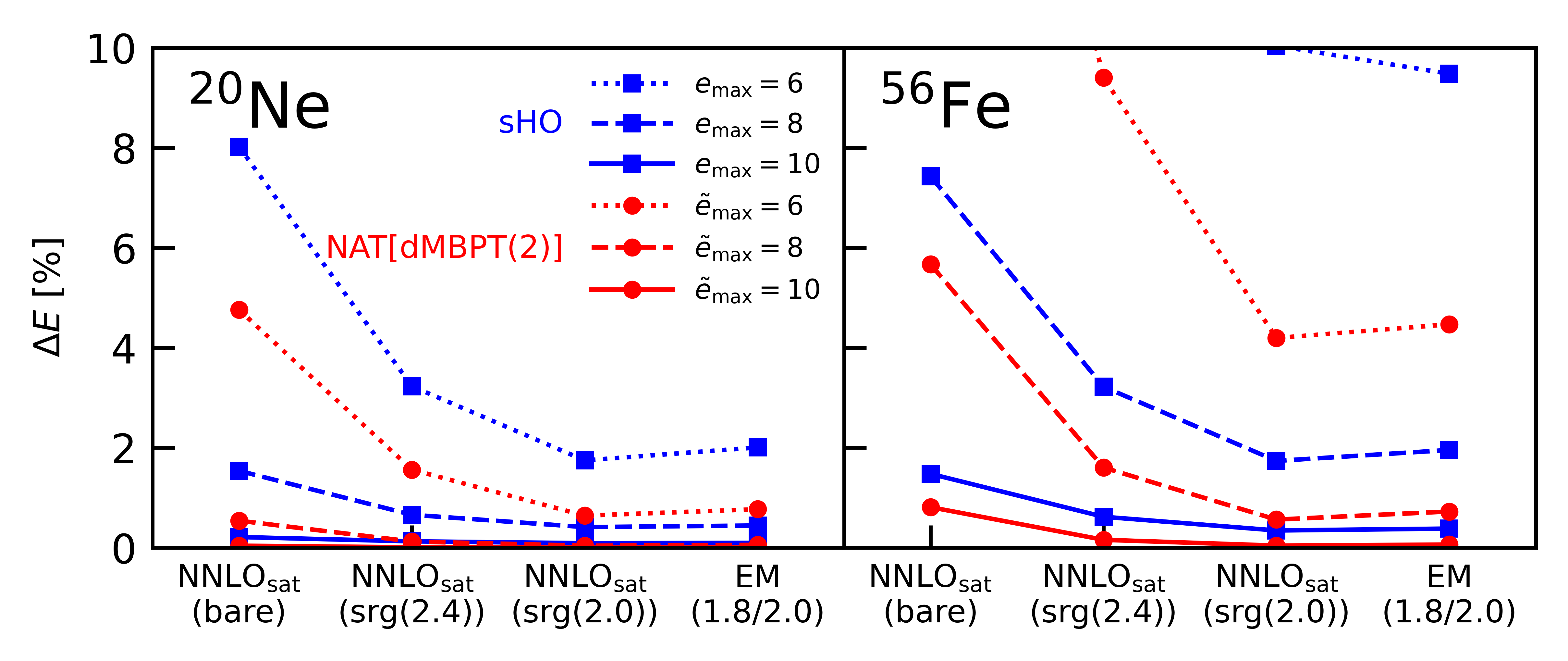}
\caption{\label{fig:srg}
Relative error on the dBMBPT(2) ground-state energy of $^{20}$Ne (left) and $^{56}$Fe (right) using the sHO (NAT[dMBPT(2)]) basis for different $e_{\text{max}}$ ($\tilde{e}_{\text{max}}$) truncations and the four different $\chi$EFT Hamiltonians characterized by different resolution scales.}
\end{figure*}

Figure~\ref{fig:srg} shows the relative error on the dBMBPT(2) ground-state energy of $^{20}$Ne and $^{56}$Fe for different $e_\text{max}$ ($\tilde{e}_\text{max}$) truncations on the sHO (NAT[dMBPT(2)]) basis. The overall behaviour is similar in the two nuclei, i.e. while the error for a given $e_\text{max}$ ($\tilde{e}_\text{max}$) decreases with the resolution scale of the Hamiltonian, the relative gain offered by the NAT basis over the sHO basis is essentially independent of it. For Hamiltonians characterised by a low resolution scale, the use of the NAT[dBMBPT(2)] basis allows a 1\% error at $\tilde{e}_\text{max}=6$ ($\tilde{e}_\text{max}=8$) in $^{20}$Ne ($^{56}$Fe) while the sHO basis necessitates two more major shells to reach the same result. For the ${\rm NNLO}_{\rm sat}$ (bare) Hamiltonian characterized by the highest resolution scale, $\tilde{e}_\text{max}=8$ yields a 1\% error in $^{20}$Ne whereas two more major shells are necessary for the sHO basis. In $^{56}$Fe, however, the  NAT[dBMBPT(2)] basis does not offer a significant gain over the sHO basis when targeting a 1\% error\footnote{Since the relative error is here computed with respect to an $\emax=12$ calculation, it is worth specifying that the converged character of the ground-state energy at $\emax=12$ is rather similar for the four interactions under consideration. As a matter of fact, computing the error in Fig.~\ref{fig:srg} with respect to the extrapolated values for each interaction rather than with respect to the corresponding $\emax=12$ calculation barely changes the curves on the figure and leaves the conclusions unchanged.}.

\subsection{$\hbar\omega$ dependence}

In Ref.~\cite{Hoppe21} the independence of the results obtained in closed-shell nuclei using the NAT basis generated via spherical MBPT(2) on the $\hbar\omega$ frequency of the underlying sHO basis was highlighted. An analogous study is now carried out in deformed nuclei based on the NAT[dBMBPT(2)] basis.

\begin{figure}
\centering
\includegraphics[width=1.\columnwidth]{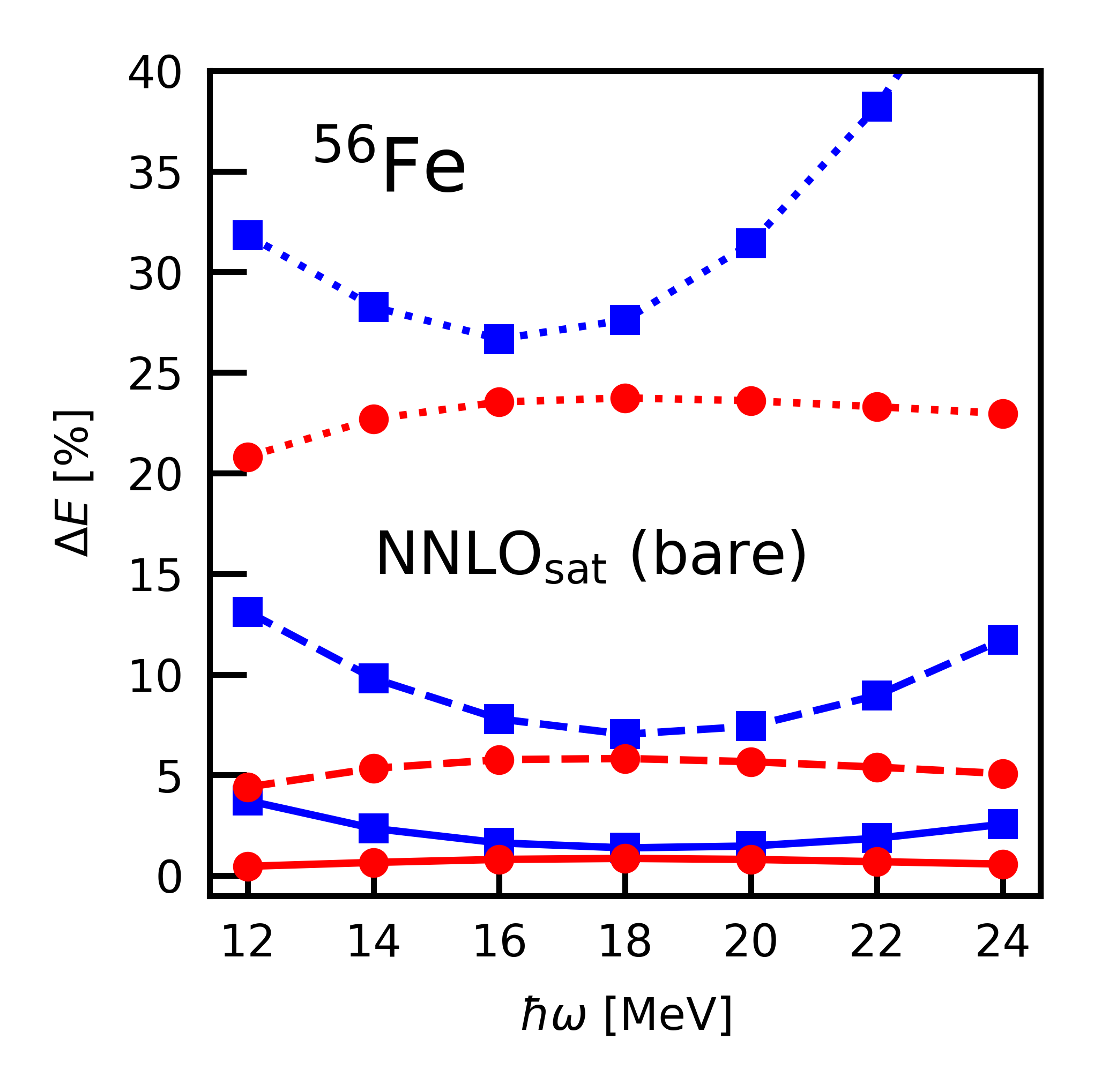}
\caption{\label{Fe_hw}
Relative error on the dBMBPT(2) ground-state energy of $^{56}$Fe as a function of the oscillator frequency $\hbar\omega$ of the underlying sHO basis. Results are shown for the ${\rm NNLO}_{\rm sat}$ (bare) Hamiltonian. The legend is the same as in Fig.~\ref{fig:srg}.
}
\end{figure}

Figure~\ref{Fe_hw} shows the relative error on the dBMBPT(2) ground-state energy of $^{56}$Fe for the ${\rm NNLO}_{\rm sat}$ (bare) Hamiltonian\footnote{As discussed in the previous section, this is the least favorable situation regarding the actual benefit of the NAT[dBMBPT(2)] basis over the sHO one. It is unimportant here given that the goal is simply to investigate how the behavior {\it evolves} with $\hbar\omega$.} as a function of the oscillator frequency $\hbar\omega$ of the underlying sHO basis for different truncations of the model space.  In agreement with the results obtained in closed-shell nuclei~\cite{Hoppe21}, the relative error is flattened for the NAT[dBMBPT(2)] basis compared to the sHO basis. This behaviour originates from the fact that the dBMBPT(2) calculation is essentially converged at $e_\text{max}=12$ and thus $\hbar\omega$-independent, meaning the corresponding density matrix is as well\footnote{Further considerations about the $\hbar\omega$-dependence of the NAT orbitals are made in Sec.~\ref{spwfs}}.

One observes that the benefit obtained from the NAT basis is minimal for $\hbar\omega = 18$ MeV, which corresponds to the optimal frequency for ${\rm NNLO}_{\rm sat}$ (bare) as far as the convergence of the results based on the sHO basis is concerned. On the other hand, the independence of the NAT basis on $\hbar\omega$ can be used to avoid searching for such an optimal frequency and thus save significant computational resources.

\subsection{Isotopic dependence}
\label{range}

Having characterised the performance of the NAT basis for different nuclear masses, the evolution along nine even-even iron isotopes ranging from $^{40}$Fe to $^{72}$Fe is now investigated. At the same time, the impact of using one fixed NAT basis extracted from, e.g., $^{56}$Fe (i.e. the NAT[dMBPT(2), $^{56}$Fe] basis) for all the isotopes is also studied. One might indeed expect that the characteristics of the natural orbitals do not evolve significantly along an isotopic chain or even within a given mass region. If so, the CPU time needed to repeatedly perform a dBMBPT(2) calculation to extract the NAT[dMBPT(2)] basis and transform the matrix elements of all operators at play into that basis could be avoided whenever performing a systematic study.

\begin{figure}
\centering
\includegraphics[width=1.0\columnwidth]{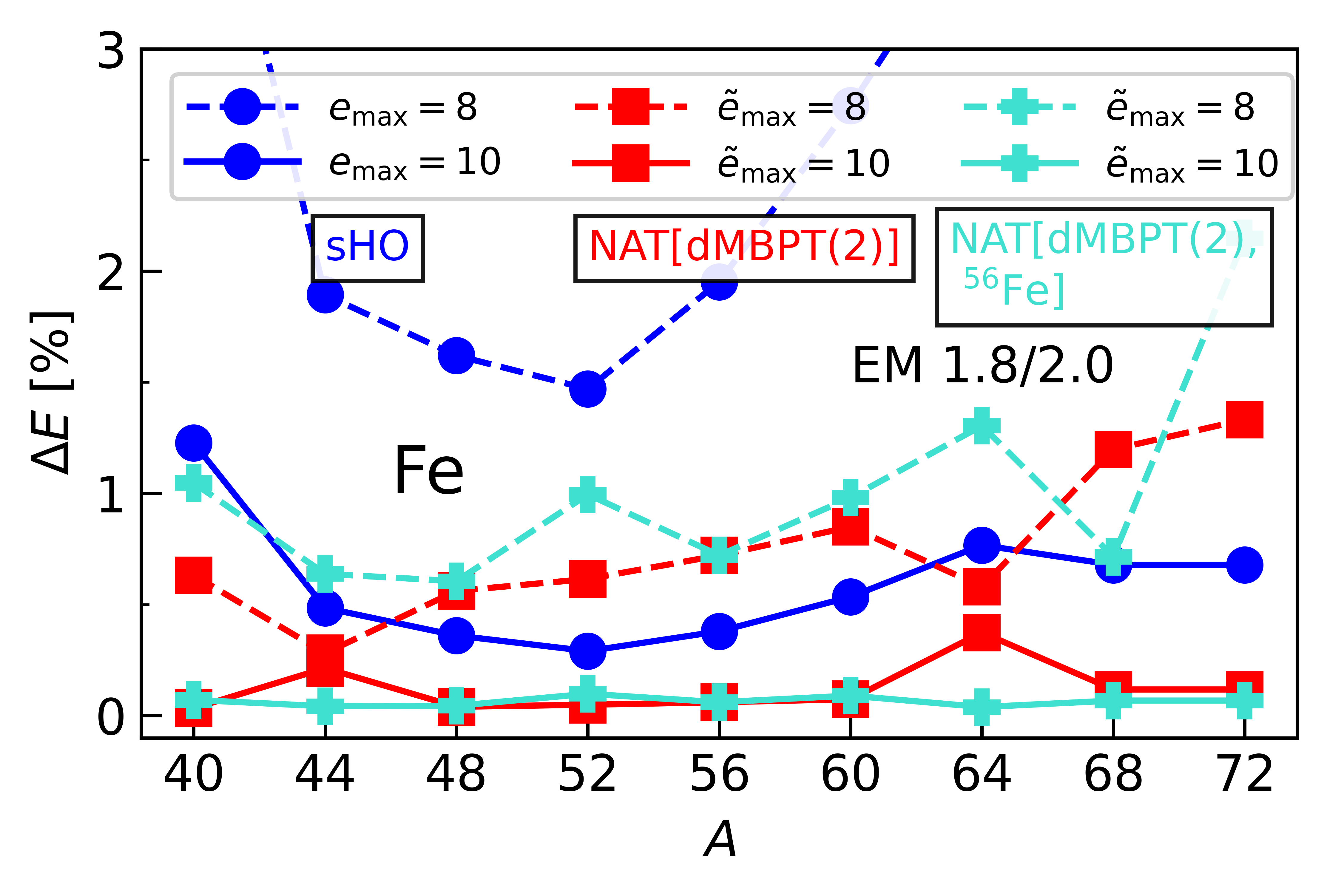}
\caption{\label{fig/Fe_chain.png} Relative error on the dBMBPT(2) ground-state energy along the Fe isotopic chain. Results are shown using both the nucleus-dependent NAT[dMBPT(2)] basis and the fixed  NAT[dMBPT(2), $^{56}$Fe] basis for all isotopes. Calculations were performed with the EM 1.8/2.0 Hamiltonian.
}
\end{figure}

The results obtained along the Fe isotopic chain with the EM 1.8/2.0 Hamiltonian are displayed in Fig.~\ref{fig/Fe_chain.png}. First, the benefit of using the NAT[dMBPT(2)] basis identified earlier for $^{56}$Fe extends similarly to all isotopes under consideration. Second, one observes that keeping the NAT[dMBPT(2), $^{56}$Fe] the same for the nine isotopes does not deteriorate the results, i.e. the gain compared to using the sHO basis remains essentially the same.  This demonstrates that using NAT orbitals computed in a nearby nucleus indeed represents a viable option. Such a study could be extended to a larger range of nuclei in the future to identify the limit of such a strategy.

\section{Alternatives to NAT[dBMBPT(2)]}
\label{investigation}

A key feature of natural orbitals relates to their capacity to carry fingerprints of correlations imprinting the many-body wave function.
This is first reflected into their optimal average occupation profile (see Sec.~\ref{propNO}), which is exploited to construct efficient truncations of the one-body basis.
One might thus wonder whether other ways\footnote{Any useful alternative must be characterised by a low computational cost to be worth considering. For instance, even though natural orbitals extracted from a more refined (and costlier) calculation than dBMBPT(2) are expected to be more efficient, following this route would defy the original purpose.} of incorporating information about the correlated wave function into the single-particle basis provide an advantage over the sHO basis.

\subsection{Alternatives}
\label{alternatives}

A first option consists in extracting the NAT basis from a deformed HFB many-body state, i.e. in using the so-called canonical basis from HFB theory~\cite{RiSc80}. Because the canonical basis is the NAT basis of a many-body state capturing static pairing correlations, canonical states are indeed known to be all localized~\cite{Tajima:2003mc} and to decay faster than the one-body local density distribution.

Instead of diagonalising the one-body density matrix, another interesting option consists in utilising the eigenbasis of the one-body Baranger Hamiltonian~\cite{BARANGER1970225}
\begin{equation}
h^{\rm BAR}_{\alpha\beta} = t_{\alpha\beta} + \sum_{\gamma\delta}v_{\alpha\gamma\beta\delta} \, \rho_{\delta\gamma} \, ,
\label{eqn:bar}
\end{equation}
where $t_{\alpha\beta}$ and $v_{\alpha\gamma\beta\delta}$ denote matrix elements of the one-body kinetic energy and of the two-body interaction, respectively. The eigenstates of $h^{\rm BAR}$ deliver an alternative one-body basis informed from many-body correlations through the input one-body density matrix. The Baranger (BAR) one-body basis is obtained at a similar cost as the NAT basis given that it requires to convolute the dBMBPT(2) one-body density matrix with the two-body interaction according to Eq.~\eqref{eqn:bar} prior to diagonalising the one-body Hamiltonian $h^{\rm BAR}$. In this case, the basis states can be ordered and truncated according to the associated eigenvalues\footnote{These one-body eigenergies are meaningful effective single-particle energies~\cite{Duguet15,Soma:2024scm} and are routinely evaluated in nuclear structure calculations.} of $h^{\rm BAR}$.

\begin{figure}
\centering
\includegraphics[width=1.0\columnwidth]{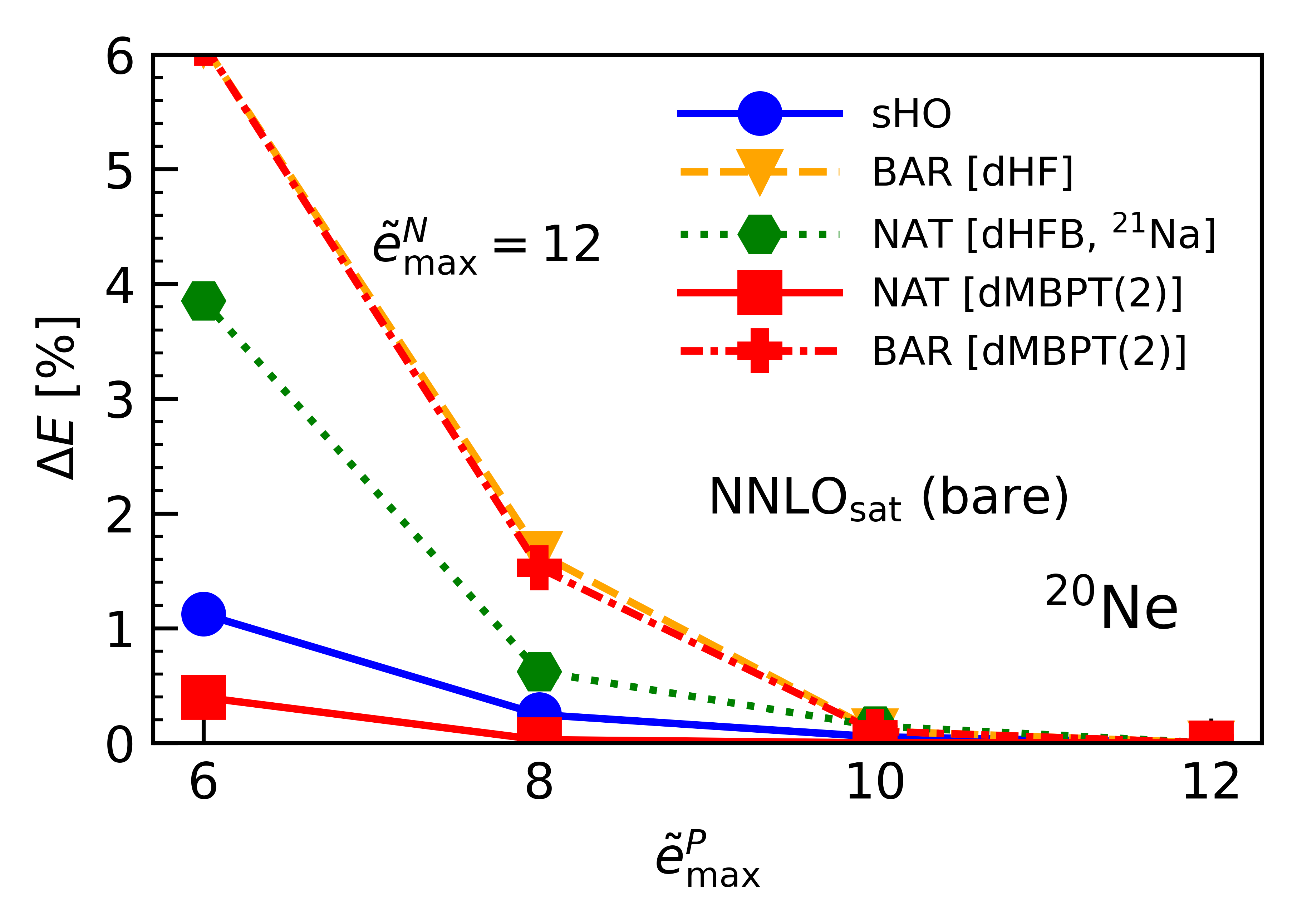}
\caption{\label{fig:Correlations} Convergence of the dBMBPT(2) ground-state energy $^{20}$Ne as a function of $\tilde{e}_\text{max}^P$ for five different one-body bases of interest (see text for details). The neutron basis is left untruncated, i.e. $\tilde{e}_\text{max}^N=12$ is used everywhere. Calculations are performed with the ${\rm NNLO}_{\rm sat}$ (bare) Hamiltonian.}
\end{figure}

\subsection{Performance}
\label{performance}

The convergence as a function of $\tilde{e}^P_\text{max}$ (keeping $\tilde{e}^N_\text{max}=12$ fixed) of the dBMBPT(2) ground-state energy obtained in $^{20}$Ne with the ${\rm NNLO}_{\rm sat}$ (bare) Hamiltonian is displayed in Fig.~\ref{fig:Correlations} for the five following proton bases 
\begin{enumerate}
    \item sHO basis;
    \item BAR[dHF] basis\footnote{The HF basis is both the NAT basis and the BAR basis associated with the HF Slater determinant. The occupations being highly degenerate (step function), such a variable does not authorise an unambiguous ordering. It is thus necessary to use Baranger (i.e. HF) single-particle energies to generate a meaningful ordering of the basis states.};
    \item NAT[dHFB, $^{21}$Na] basis obtained from the even-number parity HFB solution of the neighbouring $^{21}$Na isotone\footnote{As for the large majority of doubly open-shell nuclei computed with ab initio interactions~\cite{Scalesi24b}, the dHFB solution of $^{20}$Ne is unpaired. A simple way to enforce pairing correlations among protons is thus to use the even number-parity solution for the neigboring isotone $^{21}$Na. Given the conclusion of Sec.~\ref{range} this constitutes a well justified option.};
    \item BAR[dMBPT(2)] basis;
    \item NAT[dMBPT(2)] basis.
\end{enumerate}

\begin{figure*}
\centering
\includegraphics[width=2.1\columnwidth]{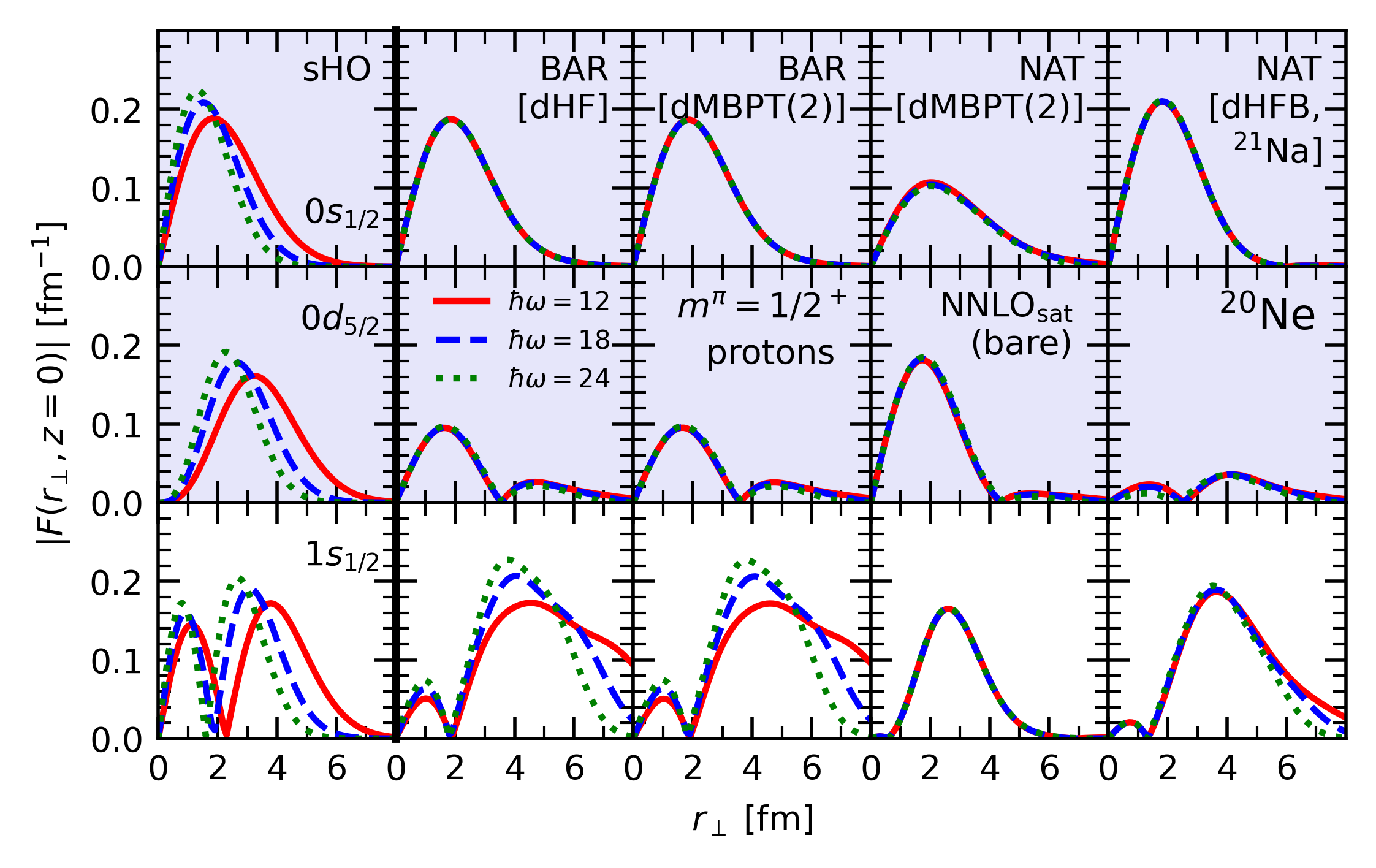}
\caption{\label{fig:wf_tot}
First three proton single-particle wave functions with $(m_\alpha=1/2,\pi_\alpha=+)$ in $^{20}$Ne as a function of $r_\perp$ for $z=0$. First column: sHO basis including the associated $(nlj)$ quantum numbers. Second column: BAR[dHF] basis. Third column: BAR[dBMBPT(2)] basis. Fourth column: NAT[dBMBPT(2)] basis. Fifth column: NAT[dHFB, $^{21}$Na] basis. The ordering of the states from top to bottom in the NAT (BAR) bases is made according to their decreasing (increasing) average occupations (Baranger single-particle energies). One-body states that would be occupied, i.e. below the Fermi level, according to a naive filling of the shells are indicated with a grey background. 
Calculations are performed with the ${\rm NNLO}_{\rm sat}$ (bare) Hamiltonian for three values of sHO basis frequency $\hbar\omega=12,18,24$ MeV.
}
\end{figure*}

First, one can appreciate the clear supremacy of the NAT[dMBPT(2)] basis, which is in fact the only one performing better than the sHO basis by typically gaining two units of $\tilde{e}_\text{max}$ over it. 

Incorporating mean-field pairing correlations into the one-body density matrix does improve over the BAR[dHF] basis but is only superior to the sHO basis for $\tilde{e}^P_\text{max}=2$ (not visible on the plot), which is irrelevant given that the error is of the order of $20-30\%$ for such small bases. This already shows that the spatial localization of the orbitals induced by pairing correlations is beneficial but not refined enough. 

The BAR[dMBPT(2)] basis and the BAR[dHF] basis display identical behaviors and provide the worst performance of all. In particular, they deliver a much slower convergence than the sHO basis. Convoluting the correlated dMBPT(2) one-body density matrix with the two-body interaction to produce and diagonalize the Baranger one-body Hamiltonian washes out the relevant fingerprint of beyond-mean-field correlations built into that density matrix. 

\subsection{Single-particle wave functions}
\label{spwfs}

To better understand the behaviour of the different one-body bases  employed in Fig.~\ref{fig:Correlations}, spatial properties of the associated wave functions are now investigated. The coordinate representation of single-particle wave functions with axial symmetry, $z$ being the coordinate along the symmetry axis and $r_\perp$ the coordinate perpendicular to it, is detailed in \ref{spwavefunctions}.

Figure~\ref{fig:wf_tot} displays in each basis, for three values of sHO basis frequency $\hbar\omega=12,18,24$ MeV, three representative proton single-particle wave functions, i.e. the first three proton states with $(m_\alpha=1/2,\pi_\alpha=+)$, as a function of $r_\perp$ (fixing $z=0$). In NAT bases, the ordering of the states from top to bottom is made according to their decreasing average occupations. In BAR bases, this ordering relates to their increasing Baranger single-particle energies. States that would be occupied in $^{20}$Ne, i.e. below the Fermi level, according to a naive filling of the shells are indicated with a grey background.

Several considerations can be made by inspecting Fig.~\ref{fig:wf_tot}
\begin{itemize}
    \item \emph{$\hbar\omega$ dependence} While the three sHO wave functions display (by construction) a dependence on the underlying sHO frequency, states {\it below} the Fermi level are independent of $\hbar\omega$ for the four other bases. The state {\it above} the Fermi level behaves, however, differently: while a significant $\hbar\omega$ dependence  is observed for both BAR bases, the dependence is considerably reduced for the NAT[dHFB, $^{21}$Na] state and disappears for the NAT[dMBPT(2)] one.
    \item \emph{Localisation} The spatial extension of the sHO states directly reflects the size, i.e. the frequency $\hbar\omega$, of the sHO potential. At long distances, sHO states behave as bound states decaying as Gaussian functions. In the four other bases, the spatial extension of the states below the Fermi level resembles their sHO counterpart obtained for the optimal $\hbar\omega=18$ value. Furthermore, these wave functions behave as bound-like state decaying exponentially at long distances\footnote{Although hardly visible in the linear $y$-scale of the figure, the validity of this statement has been explicitly verified. In practice of course, the exponential decay of the NAT states is only effective at intermediate distances, i.e. over a few fm beyond the classical turning point. At even longer distances, NAT states are bound to decay as Gaussian functions due to the finite dimension of the sHO basis they are expanded over. Still, the exponential decay in the vicinity of the nuclear volume is what eventually matters in the present study.}. A major difference occurs for the states located above the Fermi level, i.e. while the $\hbar\omega$-independent state in the NAT bases is localized within the volume of the nucleus and decays exponentially at long distances, the $\hbar\omega$-dependent state in the BAR bases is delocalized given that it corresponds to a positive Baranger single-particle energy\footnote{This delocalization is still artificially limited by the combination of the $\hbar\omega$ and $e_{\text{max}}$ values employed, i.e. the state would behave as a proper scattering state in the limits $\hbar\omega\rightarrow 0$ and/or $e_{\text{max}}\rightarrow \infty$.}. While the many-body correlations built into the dBMBPT(2) one-body density matrix efficiently localize all its eigenstates, the effect is lost when computing the Baranger Hamiltonian whose eigenstates with positive single-particle energy are scattering states independently of the correlations entering the one-body density matrix used to compute it. Eventually, one further observes that the state above the Fermi level is more localized in the NAT[dMBPT(2)] basis than in the NAT[dHFB, $^{21}$Na] basis.
    \item \emph{Nodes} Since both BAR and NAT states mix sHO states with different values of the principle quantum number $n_\alpha$, the number of nodes in the corresponding wave functions cannot be anticipated or easily interpreted. For instance, while the first state carries no node in the five bases, the second state displays one node in all bases but the sHO one.
\end{itemize}

\begin{figure}
\centering
\includegraphics[width=1.0\columnwidth]{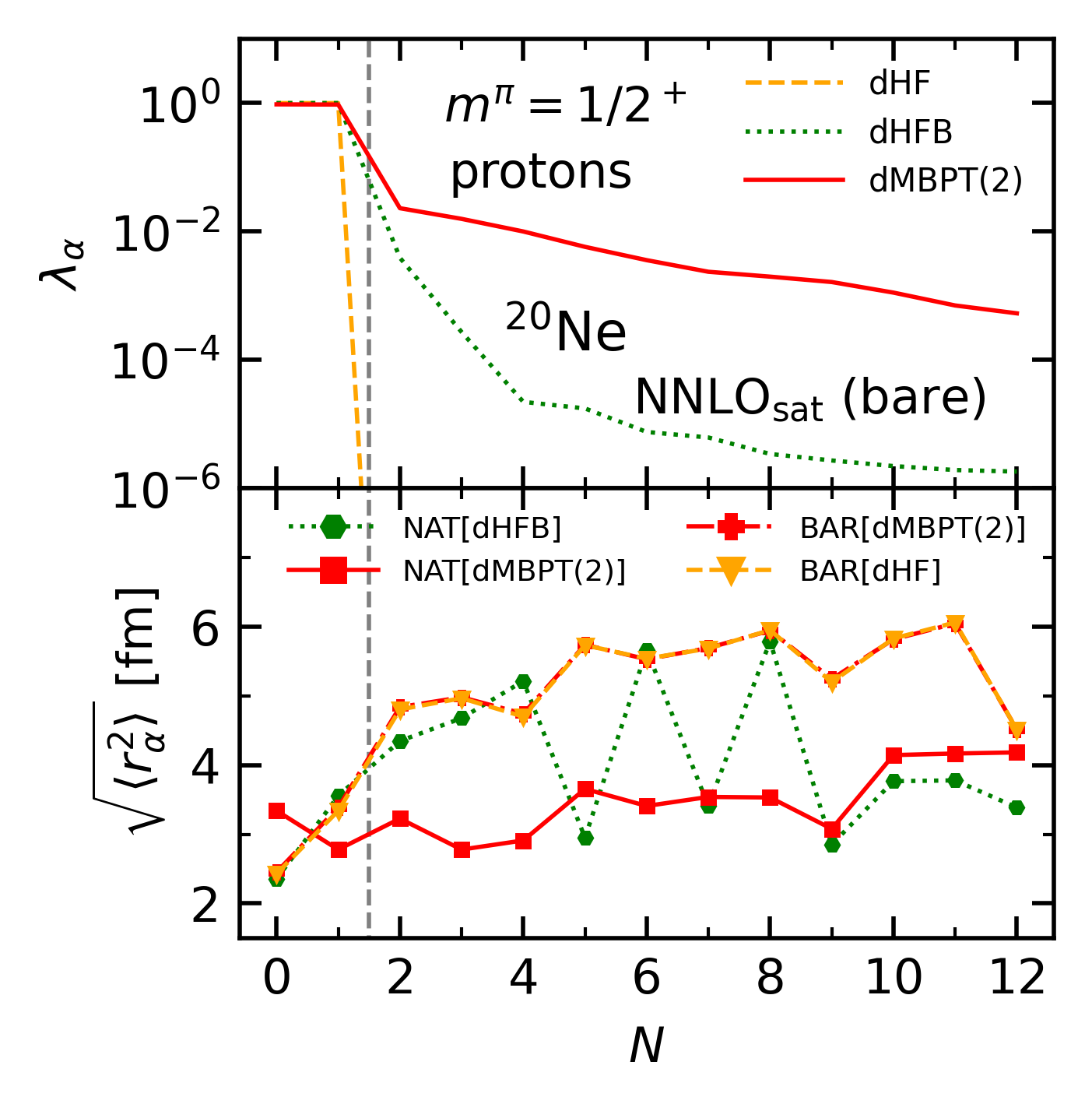}
\caption{\label{fig:rad_occ1} (Top panel) eigenvalues of the dHF, dHFB($^{21}$Na) and dMBPT(2) one-body density matrices. (Bottom panel) single-particle wave function r.m.s. radius. Results are displayed for the first states in the $[m\pi t]= [1/2+p]$ block. Calculations are performed in $^{20}$Ne with ${\rm NNLO}_{\rm sat}$ (bare) and $\hbar\omega=18$. The vertical dashed line indicates the location of the Fermi level.
}
\end{figure}

The fact that all NAT[dMBPT(2)] states are similarly localized around the volume of the nucleus is what seems to distinguish this basis from the others. To validate this conjecture, the spatial extension of the basis states is now characterised over a wider range by computing the root-mean-square (r.m.s.) radius of each basis state $\alpha$ as
\begin{align}
\sqrt{\langle r^2_\alpha \rangle} &\equiv \sqrt{\langle \alpha | r^2 | \alpha \rangle} \nonumber \\
&= 2\pi\int dr\,d\theta\,r^4\sin\theta\,F^2_\alpha(r, \theta) \, ,
\end{align}
where the function $F$ is defined in~\ref{spwavefunctions}.

Figure~\ref{fig:rad_occ1} displays the first twelve eigenvalues of the dHF, dHFB($^{21}$Na) and dMBPT(2) one-body density matrices in the $[m\pi t]= [1/2+p]$ block against the r.m.s radius of the first twelve orbitals of that same block in the BAR[dHF],  BAR[dMBPT(2)], NAT[dHFB, $^{21}$Na] and NAT[dMBPT(2)] bases. The calculation is performed in $^{20}$Ne with the ${\rm NNLO}_{\rm sat}$ (bare) Hamiltonian and the optimal frequency $\hbar\omega=18$ MeV. The following considerations can be made
\begin{itemize}
\item Even if the natural orbitals occupations are very different for the dHF and dMBPT(2) one-body density matrices, the r.m.s radii of the BAR[dHF] and BAR[dMBPT(2)] basis states are identical, i.e. the eigenfunctions of the Baranger Hamiltonian are unchanged by the correlations built into the density matrix used to compute it.
\item The spatial extension of both BAR basis states increases continuously when going from below to above the Fermi level where the r.m.s. radius of the orbitals typically reaches about $6$\,fm\footnote{The r.m.s. radius of the orbitals with positive Baranger single-particle energies would be infinite in the limits $\hbar\omega\rightarrow 0$ and/or $e_{\text{max}}\rightarrow \infty$.}. In particular, there is a large spatial mismatch between orbitals below and above the Fermi level.
\item Pairing correlations built into the dHFB($^{21}$Na) one-body density matrix only modify substantially the occupations of natural orbitals around the Fermi level such that the distribution of eigenvalues drop much faster than for dMBPT(2) natural orbitals. Eventually, the localization of the NAT[dHFB, $^{21}$Na] orbitals is not positively affected such that their r.m.s. radius remain similar to their BAR[dHF] counterparts. The calculation was repeated by boosting pairing correlations~\cite{Duguet:2020hdm} to match the occupation profile displayed by the NAT[dBMBPT(2)] orbitals in Fig.~\ref{fig:rad_occ1}. The localization of the corresponding NAT[dHFB, $^{21}$Na] orbitals was not at all improved and the convergence of the dBMBPT(2) energy was by far the worst of all tested bases.
\item Dynamical correlations built into the dMBPT(2) density matrix impact substantially the occupation profile of all natural orbitals. Eventually, the spatial extension of the NAT[dMBPT(2)] orbitals is more homogeneous than for the other bases; the r.m.s. radius typically remains between $3$ and $4$\,fm for all of them. Noticeably, the first, i.e. most occupied, NAT[dMBPT(2)] state below the Fermi level is {\it more} extended than its counterparts in the other bases (see the first row of Fig.~\ref{fig:wf_tot}) such that its spatial extension is eventually more similar to states located above the Fermi level.
\end{itemize}

\begin{figure}
\centering
\includegraphics[width=1\columnwidth]{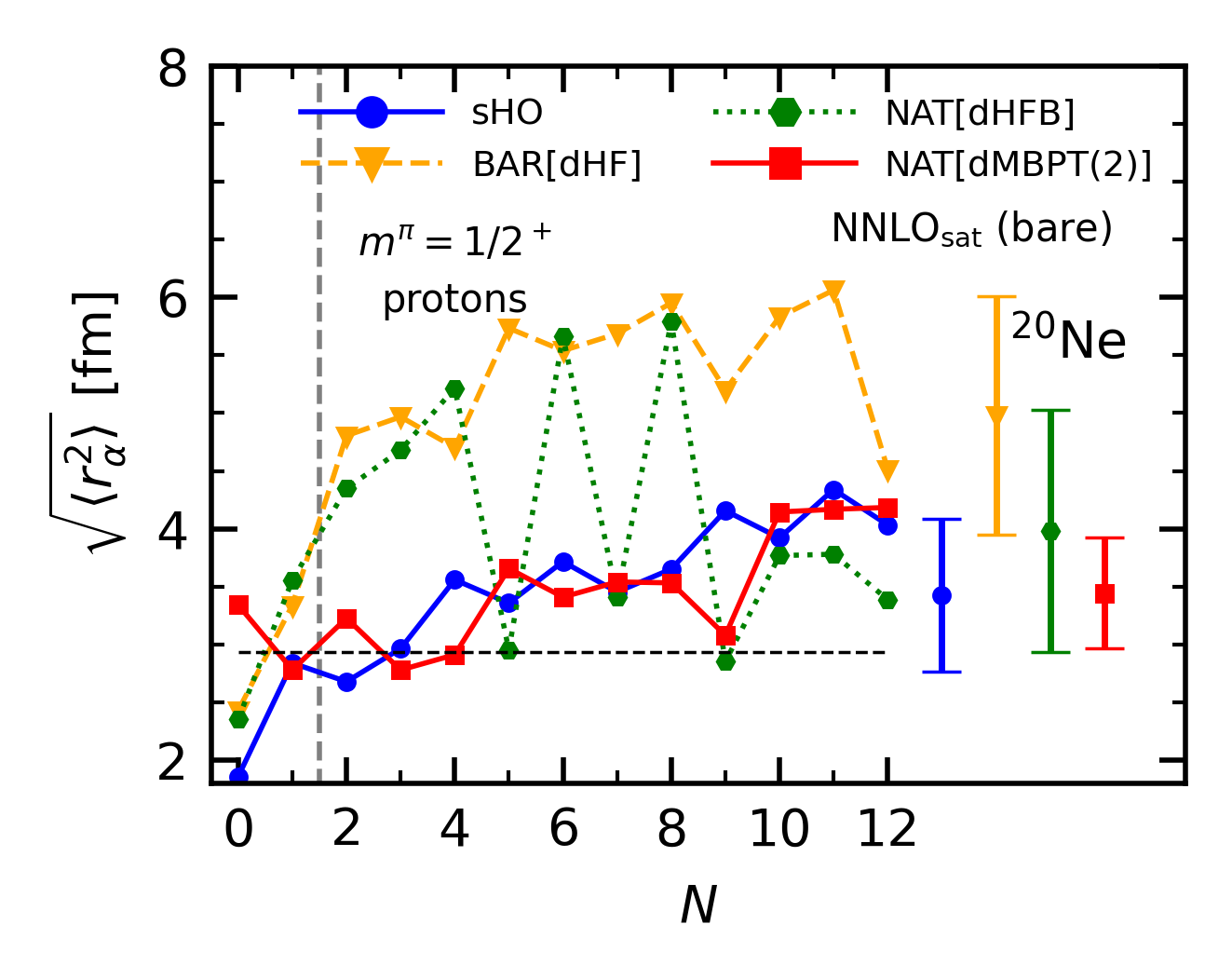}
\caption{\label{fig:rad_occ2}
Root mean square radius of the first twelve orbitals in the $[m\pi t]= [1/2+p]$ block in the sHO, BAR[dHF], NAT[dHFB, $^{21}$Na] and NAT[dMBPT(2)] bases. The average and standard deviation of the r.m.s. over the twelve states are indicated for each of the four bases. Calculations are performed in $^{20}$Ne with  ${\rm NNLO}_{\rm sat}$ (bare) and $\hbar\omega=18$. The vertical dashed line indicate the location of the Fermi level.
}
\end{figure}

Figure~\ref{fig:rad_occ2} compares in $^{20}$Ne the r.m.s. radius of the first twelve orbitals in the $[m\pi t]= [1/2+p]$ block of the sHO, BAR[dHF], NAT[dHFB, $^{21}$Na] and NAT[dMBPT(2)] bases\footnote{Results for the BAR[dMBPT(2)] basis are not shown because they are identical to those obtained with the BAR[dHF] basis.}, along with their average standard deviation, to the dMBPT(2) r.m.s. matter radius. On average, the extension of the sHO and NAT[dMBPT(2)] orbitals are more consistent with the matter radius than for the BAR[dHF] and NAT[dHFB, $^{21}$Na] basis states. Furthermore, the dispersion in the orbitals extension is the smallest for the NAT[dMBPT(2)] basis. As a matter of fact, the hierarchy in the performance of the five bases displayed in Fig.~\ref{fig:Correlations} correlates with these two spatial characteristics.

Eventually, it can be speculated that the capacity of the NAT[dMBPT(2)] basis to best converge a subsequent beyond mean-field, e.g. dBMBPT(2), calculation is correlated with the optimal spatial overlap between single-particle wave-functions below and above the Fermi level, which in turn concentrates the strength of the interaction matrix elements over the lowest lying elementary excitations. This eventually allows one to optimally build up many-body correlations as a function of $\tilde{e}_{\text{max}}$ on top of the unperturbed reference state. 

Unfortunately, natural orbitals obtained via an even less costly (pair-boosted) HFB calculation do not display appropriate properties and do not lead to any gain over the sHO basis. For a reason that remains to be elucidated, dynamical correlations brought by second order perturbation theory and static correlations brought by (boosted) HFB can lead to essentially identical eigenvalues of the one-body density matrix (i.e. natural orbitals occupation profile), while delivering very different eigenstates (i.e. natural orbital wave functions).

\section{Natural basis vs importance truncation}
\label{nat-it}

\begin{figure*}
\centering
\subfloat{{\includegraphics[width=1.00\columnwidth]{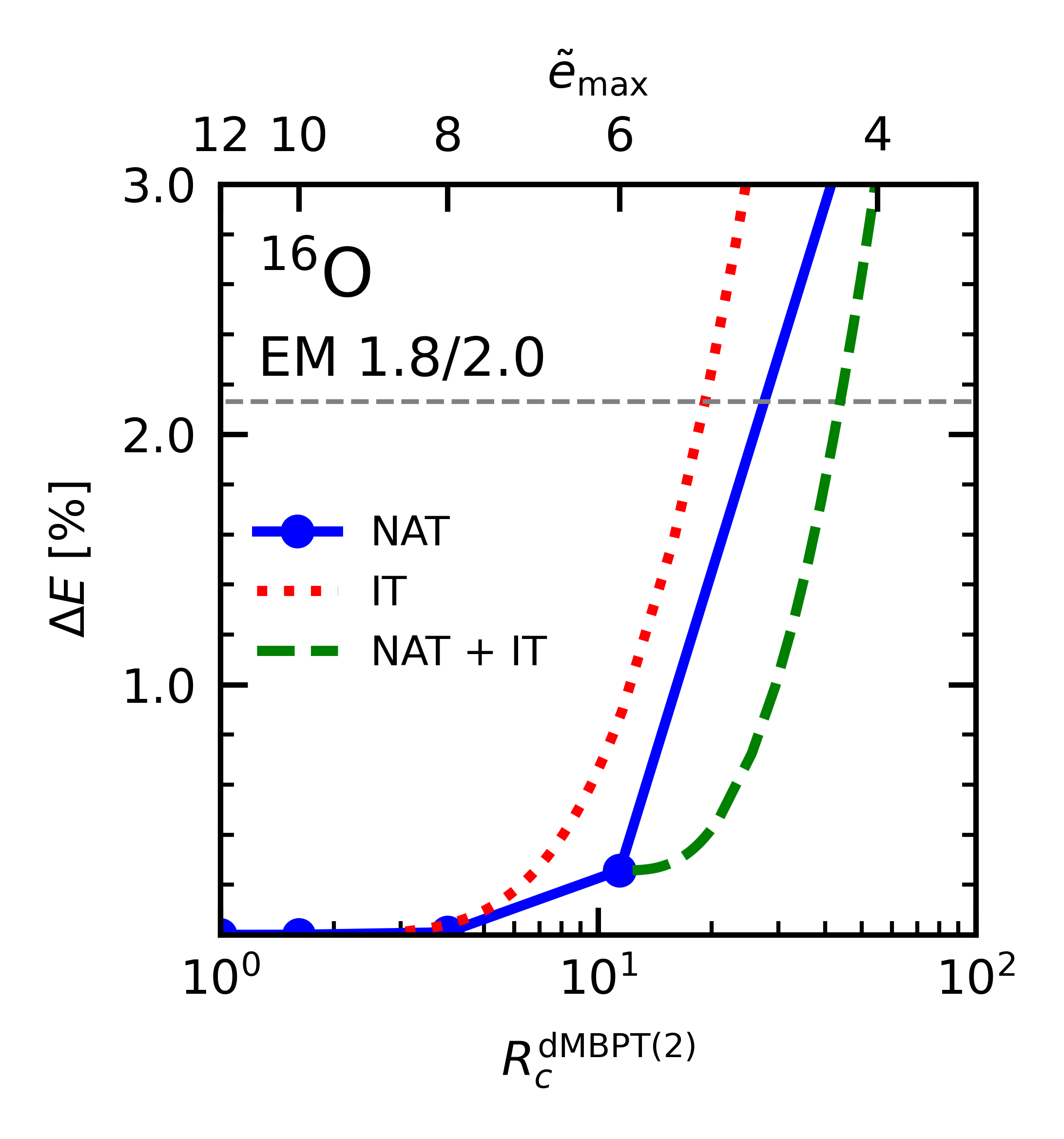} }}
%\qquad
\subfloat{{\includegraphics[width=1.00\columnwidth]{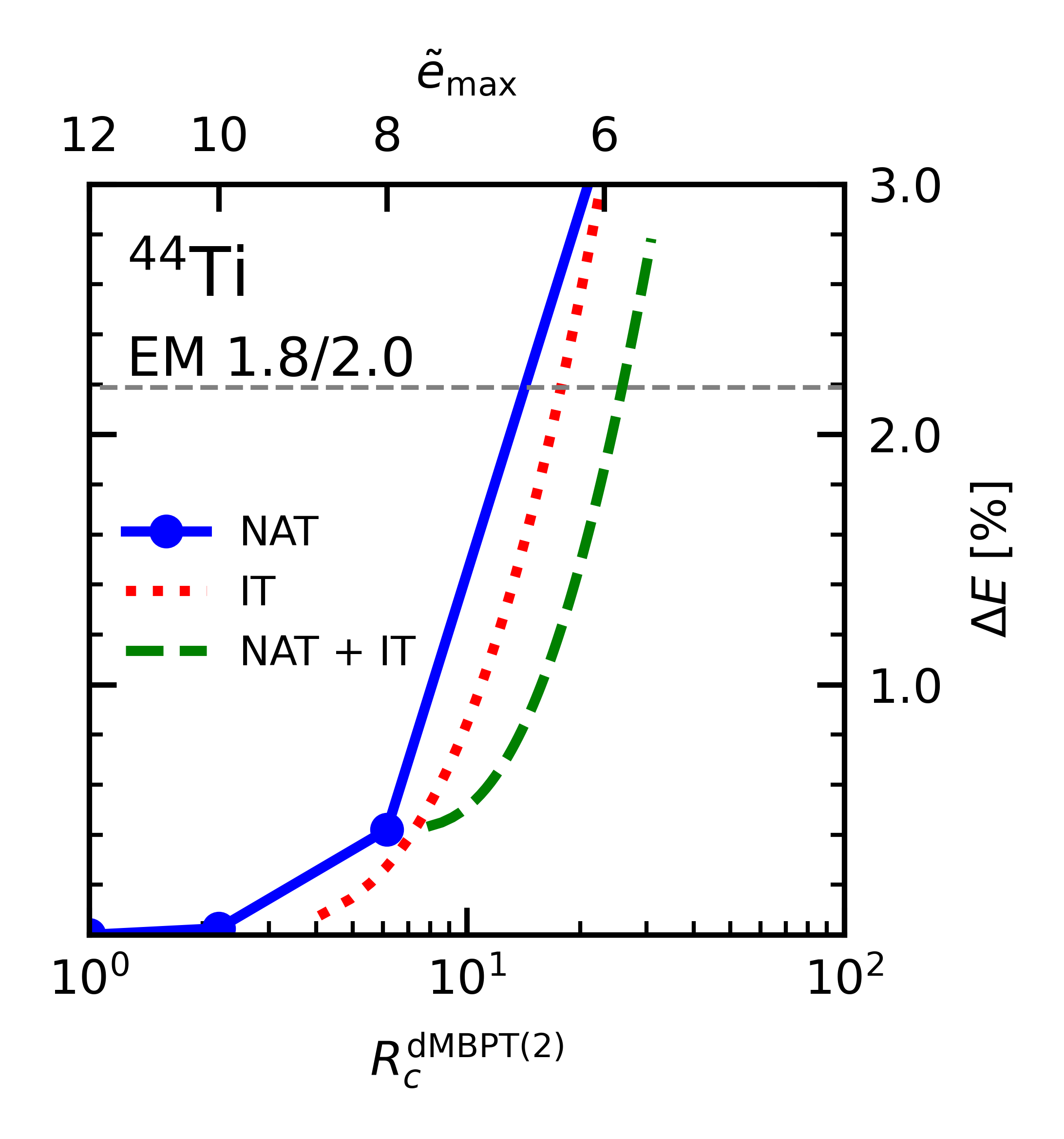} }}
\caption{Relative error on the dBMBPT(2) energy as a function of the compression factor $R_c$ using the NAT[dBMBPT(2)] basis, the IT technique and combining both. (Left panel) $^{16}$O. (Right panel) $^{44}$Ti. The dashed grey line represents the relative dMBPT(3) contribution to the total energy with respect to dMBPT(2). Calculations were performed with the EM 1.8/2.0 Hamiltonian.}
\label{fig:IT_O16_Ti44}
\end{figure*}

\begin{figure}
\centering
\includegraphics[width=1\columnwidth]{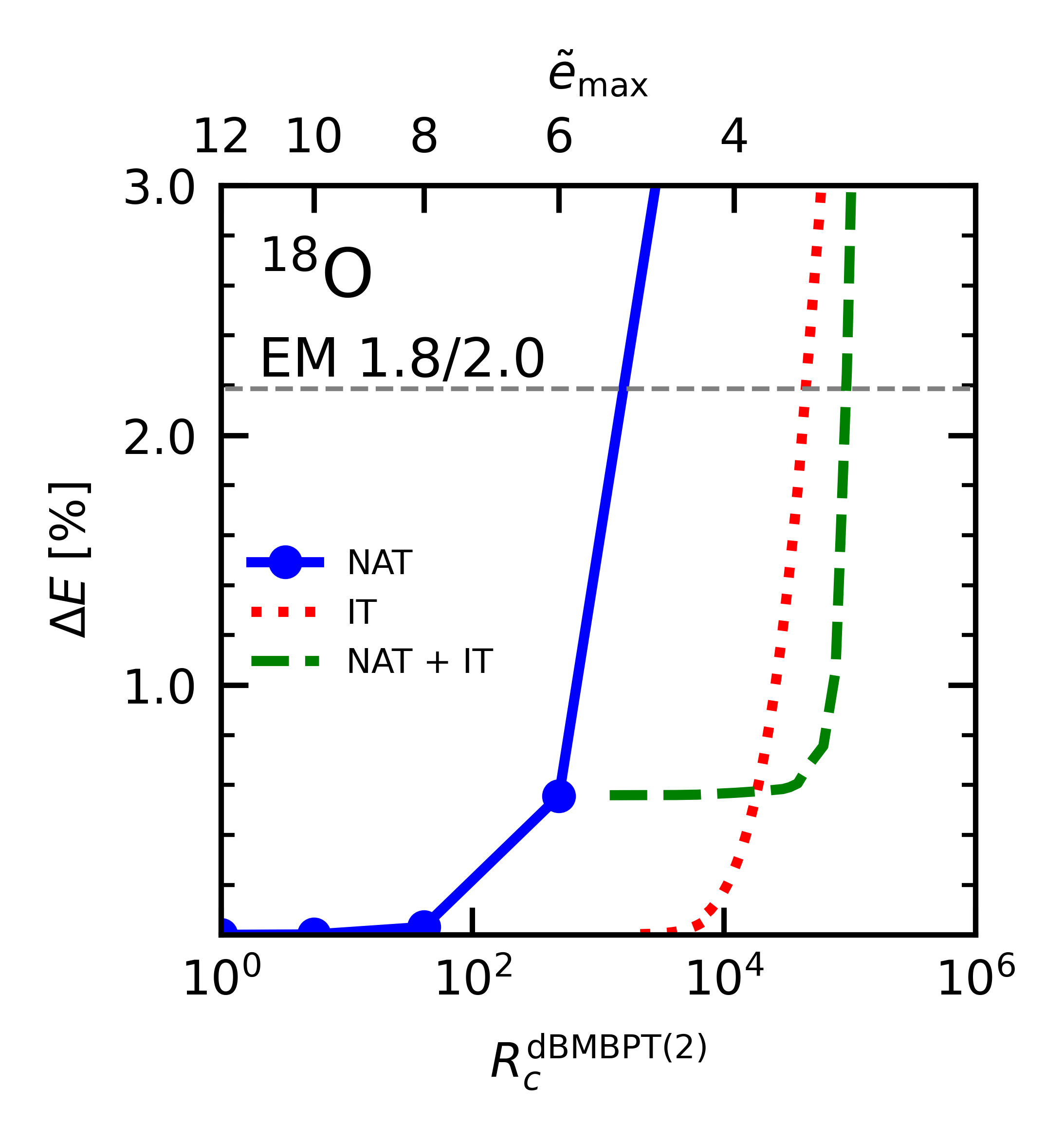}
\caption{\label{fig:IT_O18}
Same as Fig.~\ref{fig:IT_O16_Ti44} for $^{18}$O.
}
\end{figure}

\subsection{Importance truncation}

Importance truncation (IT) constitutes another well-established technique to reduce the computational costs of nuclear structure calculations while maintaining the desired accuracy on the solution of the Schr\"odinger equation.
The main idea is to pre-select, via an inexpensive evaluation, the most relevant elements of the many-body tensors at play in a method of interest. Using for example (B)MBPT(2) as the inexpensive pre-processing method, the second-order correction to the energy can be expressed as the sum over all entries of a mode-4 tensor
\begin{equation}
E^{(2)} = \frac{1}{4!} \sum_{\alpha\beta\gamma\delta} e^{(2)}_{\alpha\beta\gamma\delta}  \, , \label{BMBPT2energy}
\end{equation}
such that all quadruplets $(\alpha,\beta,\gamma,\delta)$ corresponding to entries falling below a chosen threshold $e_{\rm IT}$, 
\begin{equation}
e^{(2)}_{\alpha\beta\gamma\delta} < e_{\rm IT} \, , \label{threshold}
\end{equation}
will be ignored in a subsequent calculation involving (a counterpart of) the mode-4 tensor, the goal being to reduce the cost of expensive diagonalisations/iterations at play in non-perturbative many-body methods.
Following this strategy, IT has been successfully applied to no-core shell model~\cite{Roth:2009eu}, self-consistent Green's functions~\cite{Porro:2021rw} and in-medium SRG~\cite{IMSRG_IT} calculations. A comparison between IT and tensor factorisation techniques has also been performed within the framework of BMBPT~\cite{TICHAI2019}.

\subsection{Compression factor}

To confront the respective computational gains provided by IT and natural orbitals, using (B)MBPT(2) as the validation method, the \emph{compression factors} 
\begin{subequations}
\begin{align}
&R^{\text{d(B)MBPT(2)}}_c(e_{\rm IT}) \equiv \dfrac{n_{\text{conf}}(e_\text{max}=12, e_{\rm IT}=0)}{n_{\text{conf}}(e_\text{max}=12, e_{\rm IT})} \, ,
\label{eqn:RcBMBPT_IT}
\\
&R^{\text{d(B)MBPT(2)}}_c(\tilde{e}_\text{max}) \equiv \dfrac{n_{\text{conf}}(e_\text{max}=12, e_{\rm IT}=0)}{n_{\text{conf}}(\tilde{e}_\text{max}, e_{\rm IT} = 0)} \, ,
\label{eqn:RcBMBPT_NAT}
\end{align}
\end{subequations}
obtained with respect to a d(B)MBPT(2) calculation in $e_\text{max}=12$ and $e_{\rm IT}=0)$ are introduced. 
Such a ratio quantifies the gain by comparing the number of initial tensor entries $n_{\text{conf}}(e_\text{max}=12, e_{\rm IT}=0)$ with the number of retained tensor entries: the larger the compression factor, the greater the advantage brought by the method. The compression factor associated with the NAT basis (IT) is driven by the value of $\tilde{e}_\text{max}$ ($e_{\rm IT}$). Eventually, the number of retained entries in IT can also be translated into an effective $\tilde{e}_\text{max}$ value.

\subsection{Comparison}

Figure~\ref{fig:IT_O16_Ti44} displays the relative error $\Delta E$ on the dMBPT(2) ground-state energy against the compression factor for NAT and IT in the doubly closed-shell (open-shell) $^{16}$O ($^{44}$Ti) nucleus\footnote{The evaluation of the compression factor takes explicitly into account the fact that $U(1)$ symmetry is not broken for these two nuclei, i.e. that one can work with dMBPT(2) rather than dBMBPT(2) to begin with.}. As a rule of thumb, an acceptable error in a MBPT(2) calculation is provided by the third-order contribution appearing as a horizontal dashed line in the figure. 

In the limits $\tilde{e}_\text{max} \rightarrow 12$ or $e_{\rm IT} \rightarrow 0$, i.e. $R^{\text{dMBPT(2)}}_c=1$, the reference calculation is recovered. As $R^{\text{dMBPT(2)}}_c$ increases, the error $\Delta E$ evolves similarly for both approximation methods, even though the benefit obtained using the NAT[dMBPT(2)] basis is slightly superior (inferior) in $^{16}$O ($^{44}$Ti). Eventually, an acceptable error of $\Delta E\in [1,2]\%$ authorizes to compress the tensor at play by about one order of magnitude. Of course, the gain in non-perturbative methods pushed to high accuracy and involving mode-6 tensors to be repeatedly computed, stored and contracted is expected to be significantly higher.

\begin{figure*}
\centering
\subfloat{{\includegraphics[width=1.00\columnwidth]{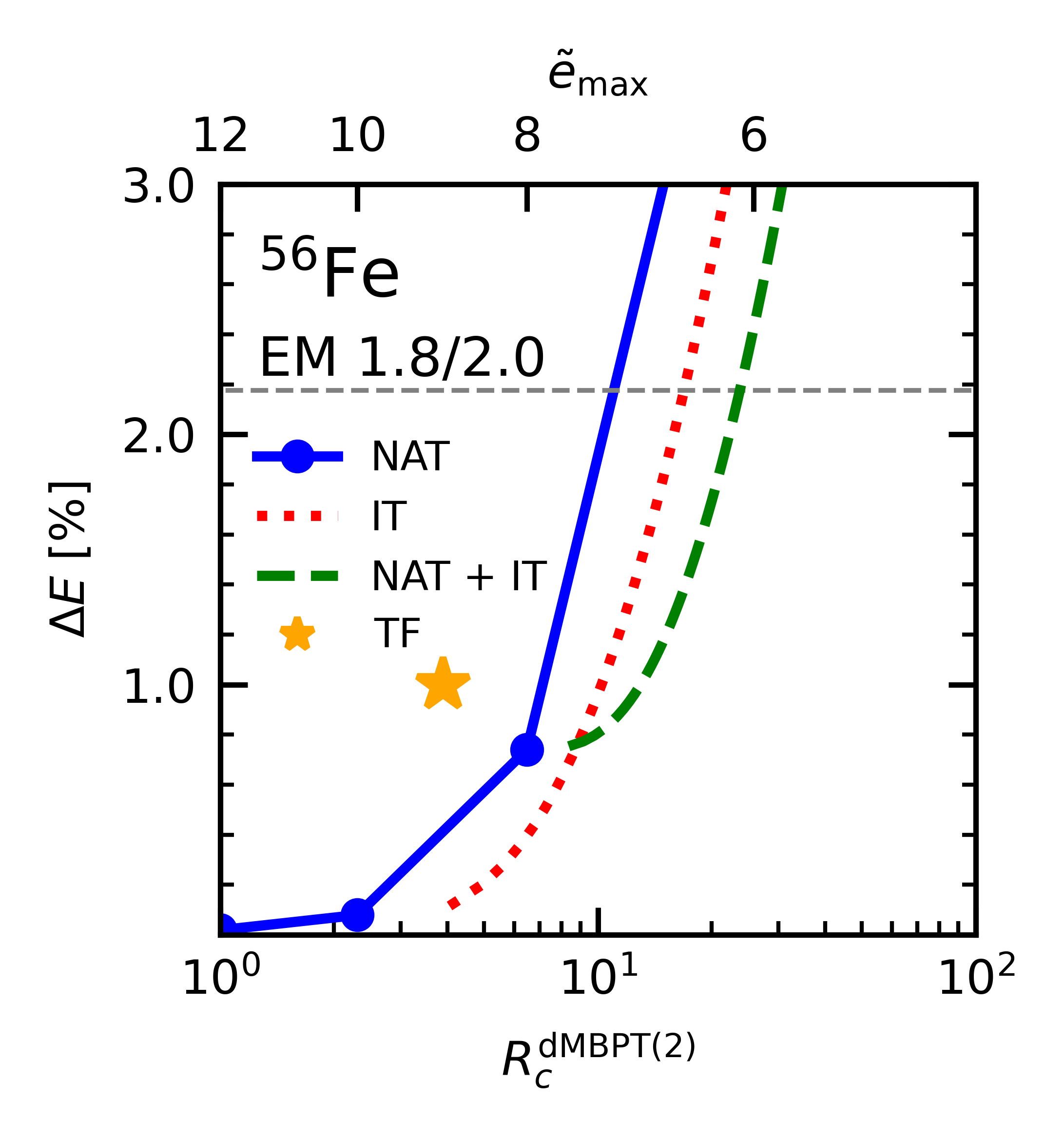} }}
%\qquad
\subfloat{{\includegraphics[width=1.00\columnwidth]{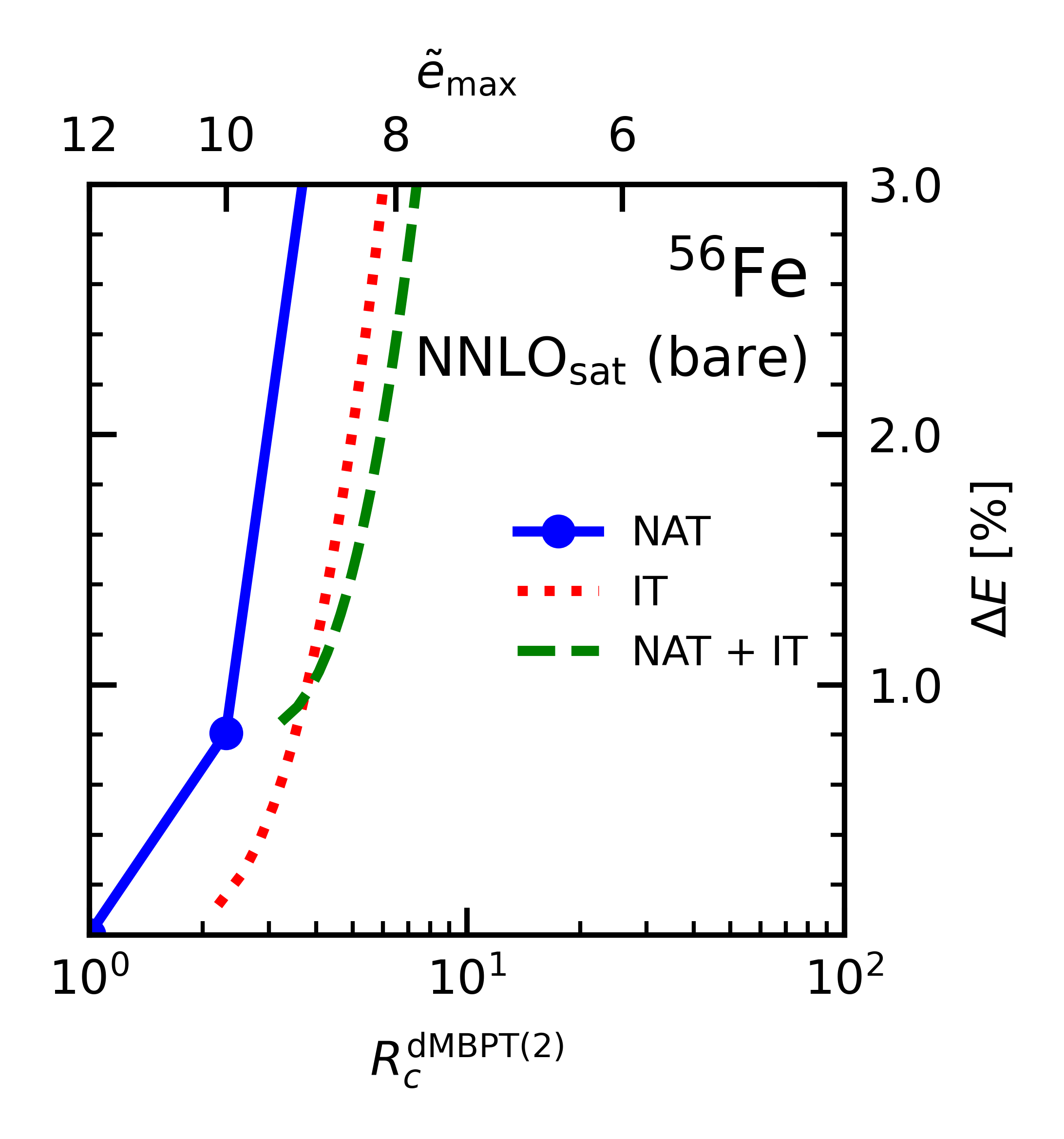} }}
\caption{Same as Fig.~\ref{fig:IT_O16_Ti44} for $^{56}$Fe only and the EM 1.8/2.0 (left panel) and  ${\rm NNLO}_{\rm sat}$ (bare) (right panel) Hamiltonians. The compression factor obtained for $\Delta E = 1.0\%$ and $e_{\text{max}}=12$ via tensor factorization techniques~\cite{frosini2024tensor} is also indicated on the left panel.}
\label{fig:IT_Fe56_mag_sat}
\end{figure*}

The situation is different in the Bogoliubov setting, as shown in Fig.~\ref{fig:IT_O18} for $^{18}$O. Indeed, the necessity to rely on the Bogoliubov algebra enlarges significantly the size of the tensors at play in the reference calculation to begin with. Both compression  techniques counterbalance this increase through larger compression factors than in the MBPT(2) case.
While reaching an error of the order of the third-order contribution ($\Delta E \approx 2\%$) via NAT generates a compression factor of $R^{\text{dMBPT(2)}}_c \approx 10^3$, IT manages to do so while compressing the tensor by one more order of magnitude.
As already observed in Ref.~\cite{Porro:2021rw} for IT and in Ref.~\cite{frosini2024tensor} for tensor factorization, Fig.~\ref{fig:IT_O18} demonstrates that the large overhead induced by the explicit treatment of pairing correlations is to a large extent artificial and can be alleviated via different pre-processing techniques.

In order to gauge how compression factors vary with the resolution scale of the input Hamiltonian,  Fig.~\ref{fig:IT_Fe56_mag_sat} compares the results obtained in $^{56}$Fe with the EM 1.8/2.0 and the ${\rm NNLO}_{\rm sat}$ (bare) Hamiltonians. While the qualitative behavior is similar, the compression factor achieved for a given error is two orders of magnitude smaller with ${\rm NNLO}_{\rm sat}$ (bare) than with EM 1.8/2.0. Although both optimisation techniques might still bring sizeable benefits for interactions characterised by higher resolution scales, much more is to be gained with soft nuclear Hamiltonians.

\subsection{Combination}

Starting from these encouraging results, NAT and IT can in fact be combined straightforwardly, i.e. the IT can be employed based on a NAT basis truncated to an appropriate $\tilde{e}_\text{max}$ value. Corresponding results are shown for one particular $\tilde{e}_\text{max}$ value\footnote{Specifically, the smallest $\tilde{e}_\text{max}$ for which $\Delta E < 1\%$ is used.} in each of the panels of Figs.~\ref{fig:IT_O16_Ti44}, \ref{fig:IT_O18} and \ref{fig:IT_Fe56_mag_sat}. In all cases, combining IT with NAT does bring a further advantage, i.e. typically a factor of 2 better than the best of the two methods used separately. When using ${\rm NNLO}_{\rm sat}$ (bare) though, the additional gain is essentially negligible.

As a final comparison, the left panel of Fig.~\ref{fig:IT_Fe56_mag_sat} also displays the compression factor obtained for $\Delta E = 1\%$ with tensor factorisation techniques~\cite{frosini2024tensor}. The compression factor is about half of the one achieved using NAT or IT in this case\footnote{It must however be noticed that the chosen example does not correspond to one for which tensor factorization provided the best benefit~\cite{frosini2024tensor}.}.

\section{Conclusions}
\label{conclusions}

The present work investigated in details the computational gain delivered from the natural basis computed via deformed second-order perturbation theory in the context of \emph{ab initio} calculations of doubly open-shell nuclei based on expansion many-body methods using an axially deformed and superfluid reference state. In view of searching for alternative bases or for natural orbitals extracted at an even lower computational cost than deformed second-order perturbation theory, the key characteristics of natural orbitals were investigated. Eventually, the use of natural orbitals was compared to the benefit brought by other compression techniques, i.e. importance truncation and tensor factorization techniques.

The main conclusions are that
\begin{itemize}
\item The natural orbital basis extracted via second-order many-body perturbation theory authorizes to converge a calculation, e.g. the same second-order many-body perturbation theory calculation,  based on soft interactions to a given accuracy using about half the number of states $n_B$ needed with the spherical harmonic oscillator basis. While the result is valid for doubly closed-shell, singly open-shell and doubly open-shell nuclei, the gain is significantly reduced when using a Hamiltonian characterized by a large resolution scale. 
\item Based on the hypothesis that such a gain extends to any many-body expansion method whose intrinsic memory load (CPU cost) scales as $n^p_B$ ($n^q_B$), the benefit of using the natural basis over the spherical harmonic oscillator basis is thus estimated to be of the order $2^p$ ($2^q$). 
\item Using a common reference calculation employing a spherical harmonic oscillator basis of given dimension (e.g. $e_{\text{max}}=12$), the gain obtained via importance truncation and tensor factorization techniques is similar to the one presently achieved based on the use of the natural orbital basis. 
\item Employing importance truncation techniques {\it on top of} a natural orbital basis allows one to gain an additional factor of $2$ in the compression of the mode-2 tensor at play in a d(B)MBPT(2) calculation compared to the benefit obtained by the best of both methods used separately.
\item While the gain characterized in the present paper is based on an $e_{\text{max}}$-like truncation of the natural orbital basis, there exists an entire freedom in the way the basis can be cut. Thus, the possibility to design more optimal truncation schemes needs to be investigated in the future.
\item None of the alternative bases presently investigated, i.e. the natural basis extracted from Hartree-Fock and (pair-boosted) Hartree-Fock-Bogoliubov calculations or the so-called Baranger basis, was shown to provide an advantage over the spherical harmonic oscillator basis. The merit of the natural basis extracted from a second-order many-body perturbation theory seems to relate to its unique capacity to localise all its orbitals over the volume occupied by the nucleus. 
\end{itemize}

\section*{Acknowledgements}

The work of A.S. was supported by the European Union’s Horizon 2020 research and innovation program under grant agreement No 800945 - NUMERICS - H2020-MSCA-COFUND-2017.
Calculations were performed by using HPC resources from GENCI-TGCC (Contracts No. A0130513012 and A0150513012).
The work of M.F. is supported by the CEA-SINET project.

\section*{Data Availability Statement}
This manuscript has no associated data or the data will not be deposited.

\appendix

\section{Single-particle wave functions in axially-deformed bases}
\label{spwavefunctions}

A generic spherical wave function defined using a so-called $j$-coupling representation

\begin{equation}
\langle r \theta \phi \sigma \tau | n j m \pi t \rangle \equiv \psi_{njm\pi t}(r, \theta, \phi, \sigma, \tau) \, ,
\end{equation}
can be represented onto a basis employing a $ls$-coupling scheme through
\begin{equation}
\begin{split}
&\psi_{njm\pi t}(r, \theta, \phi, \sigma, \tau)\\
&= \sum_{lm_lsm_s}
\psi_{nlm_l}(r, \theta, \phi)
\chi_{m_s}(\sigma)
\chi_{t}(\tau)
\langle lm_lsm_s|jm \rangle,
\label{eqn:psij}
\end{split}
\end{equation}
where $m_l$ is constrained by the sum rule in the Clebsch-Gordan coefficient ($m = m_l + m_s$), $s=1/2$ for fermions and $l$ is constrained by the knowledge of $j$ and $\pi$ ($l = |j-s|, j+s$). As a consequence, the summation in Eq.~\eqref{eqn:psij} is limited to a summation over the spin projection $m_s$. Clebsch-Gordan coefficients assume a particularly simple expression for the few values of $j$ and $m_s$ allowed, which are summarized in Tab.~\ref{tab:CG} (see Ref.~\cite{Varshalovic} for more details).

\begin{table}[h!]
\centering
\begin{tabular}{ccc}
\toprule
$j$ & $m_s = 1/2$ & $m_s = -1/2$\\
\midrule
$l+1/2$ & $\sqrt{ \dfrac{l+m+1/2}{2l+1}}$ & $\sqrt{\dfrac{l-m+1/2}{2l+1}}$\\
$l-1/2$ & $-\sqrt{ \dfrac{l-m+1/2}{2l+1} }$ & $\sqrt{ \dfrac{l+m+1/2}{2l+1} }$\\
\bottomrule
\end{tabular}
\caption{Analytical expression for the Clebsch-Gordan coefficients entering Eq.~\eqref{eqn:psij}.}
\label{tab:CG}
\end{table}
In general the $ls$-scheme wave function in Eq.~\eqref{eqn:psij} can also depend on $m_s$ and $t$ quantum numbers. However, HO wave functions are independent on the spin and ispospin projections (i.e. $\psi_{nlm_lm_st} = \psi_{nlm_l}$).
Eventually the $j$-scheme wave function can be written as a function of the two spin components ($m_s = +1/2 = \,\,\uparrow$ and $m_s = -1/2 = \,\,\downarrow$) according to
\begin{equation}
\begin{split}
&\Psi_{njm\pi t}(r, \theta, \phi, \sigma, \tau) =
\begin{pmatrix}
\Psi_{njm\pi t}(r, \theta, \phi, \sigma, \tau)_\uparrow\\
\Psi_{njm\pi t}(r, \theta, \phi, \sigma, \tau)_\downarrow
\end{pmatrix}\\
&= \sqrt{ \dfrac{1}{2} + \dfrac{2(j-l)m}{2l+1}} \psi_{nl(m-1/2)}(r, \theta, \phi)
\chi_{1/2}(\sigma)
\chi_{t}(\tau)\\
&+ (-1)^{1/2+l-j}\sqrt{ \dfrac{1}{2} - \dfrac{2(j-l)m}{2l+1}} \psi_{nl(m+1/2)}(r, \theta, \phi)\\
&\times \chi_{-1/2}(\sigma)\chi_{t}(\tau) \, ,
\label{eqn:psi}
\end{split}
\end{equation}
which is a two-dimensional vector whose components are scalar wave functions.
The wave function $\psi_{nlm_l}$ can be split into radial and angular contributions
\begin{equation}
\psi_{nlm_l}(r, \theta, \phi) = f_{nl}(r)Y_{lm_l}(\theta, \phi) \, ,
\end{equation}
where $Y_{lm_l}(\theta, \phi)$ represents a \emph{spherical harmonic}
\begin{equation}
Y_{lm_l}(\theta, \phi) \equiv \sqrt{\dfrac{2l+1}{4\pi}\dfrac{(l-m_l)!}{(l+m_l)!}}P_l^{m_l}(\cos{\theta})e^{im_l\phi}.
\end{equation}
Eventually, the dependencies on the three spatial coordinates separate as
\begin{equation}
\begin{split}
&\psi_{nlm_l}(r, \theta, \phi)\\
&= \underbrace{f_{nl}(r)}_{r} \underbrace{\sqrt{\dfrac{2l+1}{4\pi}\dfrac{(l-m_l)!}{(l+m_l)!}}P_l^{m_l}(\cos{\theta})}_{\theta}\underbrace{e^{im_l\phi}}_{\phi} \, .
\label{eqn:wftot}
\end{split}
\end{equation}
A single-particle scalar wave function in $j$-scheme is characterised by the set of quantum numbers $n$, $j$, $\pi$, $m$ and $m_s$. The radial function $f_{nl}(r)$ in Eq~\eqref{eqn:wftot} is often re-written as a function of the \emph{reduced radial wave function} $u_{nl}(r)$
\begin{equation}
f_{nl}(r) = \dfrac{u_{nl}(r)}{r}.
\end{equation}

Introducing the quantity
\begin{equation}
F^{\rm SPH}_{nlm_l}(r, \theta) = {u_{nl}(r)} {\sqrt{\dfrac{2l+1}{4\pi}\dfrac{(l-m_l)!}{(l+m_l)!}}P_l^{m_l}(\cos{\theta})} \, ,
\label{eqn:Fsph}
\end{equation}
its $j$-scheme version is given by
\begin{equation}
\begin{split}
&F^{\rm SPH}_{njm\pi t}(r, \theta, \sigma) =
\begin{pmatrix}
F^{\rm SPH}_{njm\pi t}(r, \theta, \sigma)_\uparrow\\
F^{\rm SPH}_{njm\pi t}(r, \theta, \sigma)_\downarrow
\end{pmatrix}\\
&=
\begin{pmatrix}
\sqrt{ \dfrac{1}{2} + \dfrac{2(j-l)m}{2l+1}} F^{\rm SPH}_{nlm-1/2}(r, \theta)\\
(-1)^{1/2+l-j}\sqrt{ \dfrac{1}{2} - \dfrac{2(j-l)m}{2l+1}} F^{\rm SPH}_{nlm+1/2}(r, \theta) \, .
\end{pmatrix}
% \label{eqn:psi}
\end{split}
\end{equation}

Considering the coefficient $C^{[m\pi t]}_{n j N}$ introduced in Eq.~\eqref{eqn:NATcoeff} for the change of basis between HO and NAT, such a coefficient can be used to mix HO quantum numbers $n$ and $j$ to give an expression for the deformed NAT wave functions
\begin{equation}
F^{\rm dNAT}_{Nm\pi t}(r, \theta, \sigma) \equiv \sum_{nj} F^{\rm sHO}_{njm\pi}(r, \theta, \sigma) C^{[m\pi t]}_{n j N} \, .
\end{equation}
This transformation conserves quantum numbers $m$, $\pi$ and $t$.

Eventually re-expressing ($r$, $\theta$) in terms of cylindrical coordinates
\begin{equation}
\begin{dcases}
r_{\perp} = r\sin{\theta} \, ,\\
z = r\cos{\theta} \, ,
\end{dcases}
\end{equation}
previous equations can be recast in terms of such new coordinates, i.e. $F_{Nm\pi t}(r, \theta, \sigma) \equiv  F_{Nm\pi t}(r_{\perp}, z, \sigma)$.

\begin{figure}
\centering
\includegraphics[width=1.0\columnwidth]{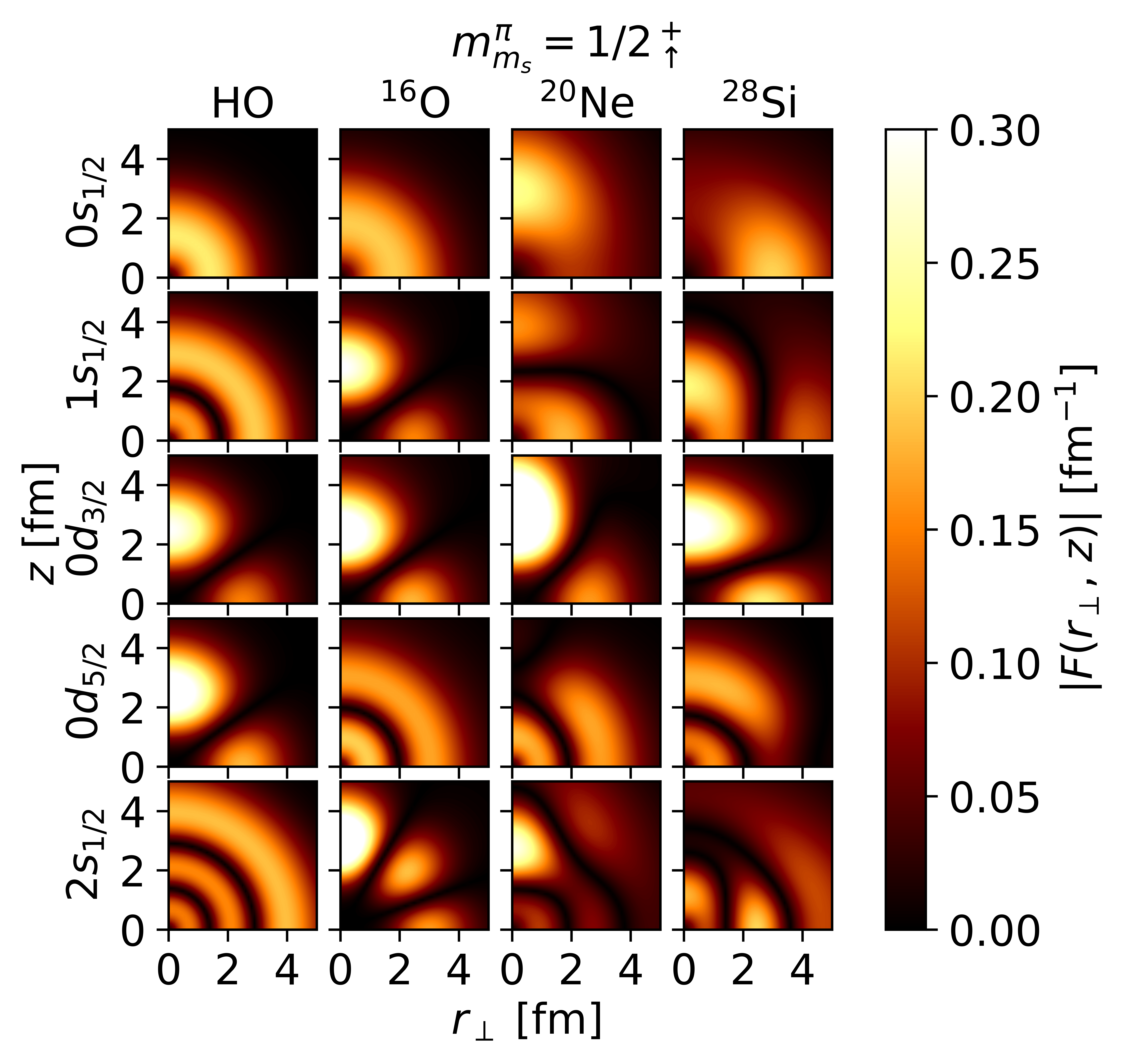}
\caption{\label{fig:WF_bidim} Two-dimensional representation of deformed orbitals in the $[m\pi t]= [1/2+p]$ block. While the first column displays sHO wave functions with $\hbar\omega= 18$ MeV corresponding to a given set of quantum numbers, the following columns corresponds to NAT[dMBPT(2)] wave functions in three different nuclei obtained using the ${\rm NNLO}_{\rm sat}$ (bare) Hamiltonian and ordered from top to bottom according to their decreasing average occupations.
}
\end{figure}

Figure~\ref{fig:WF_bidim} represents selected sHO and NAT wave functions $F(r_\perp, z)$, the latter being computed in (spherical) $^{16}$O, (prolate) $^{20}$Ne and (oblate) $^{28}$Si. 
While the $l=0$ wave functions in $^{16}$O display a symmetry along the main diagonal of the square, deformed orbitals in $^{20}$Ne and $^{28}$Si are distorted in opposite ways.

\bibliography{bibliography.bib}

\end{document}